\documentclass[prd,nofootinbib,showpacs,preprint]{revtex4}
\usepackage{graphicx}
\DeclareGraphicsExtensions{.pdf}

\usepackage{epsfig}
\usepackage{amsmath}
\usepackage{amsfonts}
\usepackage{amssymb}
\usepackage{url}
\usepackage{cancel}
\newcommand{\beqa}{\begin{eqnarray}} 
\newcommand{\eeqa}{\end{eqnarray}} 
\newcommand{\beq}{\begin{equation}} 
\newcommand{\eeq}{\end{equation}}

\newcommand{\nn}{\nonumber}
\newcommand{\bmt}{\begin{pmatrix}}
\newcommand{\emt}{\end{pmatrix}}
\usepackage[toc,page]{appendix}
\newcommand{\be}{\begin{equation}}
\newcommand{\ee}{\end{equation}}
\newcommand{\bea}{\begin{eqnarray}}
\newcommand{\eea}{\end{eqnarray}}

\begin{document}
\title{Study of the rare semileptonic  decays $B_d^0 \to K^* l^+ l^-$ in scalar leptoquark model }
\author{Suchismita Sahoo and Rukmani Mohanta }
\affiliation{School of physics, University of Hyderabad, 
              Hyderabad - 500046, India  }      
\begin{abstract}
We study the effect of scalar leptoquarks on the exclusive rare $B$ meson decays $\bar{B}_d^0 \rightarrow \bar{K}^{* 0}\left(\rightarrow K^-\pi^+\right) l^+ l^-$ 
in the full kinematically accessible physical region. We work out the constraints on leptoquark parameter space using the measured branching ratio 
of $B_s \rightarrow \mu^+ \mu^-$ process by the CMS and LHCb collaborations. We compute the branching ratio, forward-backward asymmetry 
and isospin asymmetry distribution  using the constrained  parameter space. We also look into various form factor independent and CP violating 
observables in the scalar leptoquark model.
\end{abstract}
\maketitle

\section{Introduction}

The study of rare flavour changing neutral current (FCNC) transitions of $b$-flavored mesons decaying into dileptons provide 
an ideal testing ground to critically test the standard model (SM) and to look for the possible existence of new physics (NP). 
Such processes are highly suppressed in the standard
model  as they proceed through amplitudes involving electroweak loop (penguin and box)
diagrams. Of particular importance are the rare semileptonic decays involving $b \to s \mu^+ \mu^-$ transitions,  as 
these processes are one-loop suppressed in the SM, but many extensions of the SM are capable of producing  measurable effects 
in various observables. While most of the flavor observables are in very good agreement with their  SM
predictions there are some exceptions in semileptonic $B$ decays. Recently LHCb has reported deviations from the SM expectations
in $B \to K^* \mu^+ \mu^-$  angular observables, mainly in $P_5'$ \cite{lhcb1} and decay rate \cite{lhcb-1a}, in $B_s \to \phi \mu^+ \mu^-$ decay rate
\cite{lhcb-1b} and 
in the ratio $R_K={\rm BR}(B \to K \mu^+ \mu^-)/{\rm BR}(B \to K e^+ e^-)$ \cite{lhcb-1c}. Interestingly all these deviations are
associated with the quark level transition $b \to s \mu^+ \mu^-$.  

In this paper, we would like to focus on  the
semileptonic decay mode $B_d^0 \to K^* (\to K \pi) \mu^+ \mu^- $ which is quite an 
interesting channel, as the measurement of four-body angular distribution provides a large number of
observables which can be used to probe and discriminate different scenarios of NP.
Theoretical predictions for such observables are particularly precise and free from hadronic
uncertainties in the low-range of dimuon invariant mass squared $q^2$, i.e., $1 < q^2<6~{\rm GeV}^2$.
While the observed forward-backward asymmetry is systematically below the SM prediction, the zero crossing point is
consistent with it. Also  
there are few other deviations from the SM expectations have been observed by LHCb experiment
in the angular observables. The largest discrepancy of 3.7$\sigma$ encountered in the observable  $P_5'$ \cite{lhcb1}
in the bin $q^2 \in$ [4.3, 8.68].  Another interesting observable to look for new physics is the isospin asymmetry distribution,
which is measured by LHCb experiment in the entire $q^2$ spectrum \cite{latestlhcb}. 
The leading uncertainties in the $ B\rightarrow K^*$ form factor is expected to cancel in this asymmetry.

The angular distributions of $B \rightarrow K^* l^+ l^-$  processes with the dilepton invariant mass 
has been studied by various experiments such as BaBar, Belle, CDF, 
and LHCb. All these experiments cover the full kinematical dilepton mass region, \textit{i.e.} $4m^2_l \leqslant q^2 \leqslant \left(m_B - m_{K^*}\right)^2$,
 leaving the regions around $q^2 \sim m_{J/\psi}^2$ and $m_{\psi^\prime}^2$.
In general the kinematically allowed region can be classified into three regions and different theoretical approaches usually
adopted  to study the properties of different observables.   In
 the region of large hadron recoil \emph{i.e.,}  for  $q^2\leqslant m_{J/\psi}^2$, the kaon is very energetic and
various physical observables can be computed using QCD factorization (QCDf) approach.
 The intermediate values of $q^2$ \textit{i.e.} $7~{\rm GeV}^2\leqslant q^2 \leqslant 14~{\rm GeV}^2$ fall into the narrow-resonance 
region and cuts are employed to remove the dominated charmonium resonance $\left(\bar{c}c\right) = J/\psi, \psi^\prime $ backgrounds 
from $\bar{B} \rightarrow \bar{K}^* \left(\bar{c}c\right) \rightarrow \bar{K}^* l^+ l^-$. 
The larger dilepton invariant mass region \textit{i.e.,} $q^2 \geqslant 14~ {\rm GeV}^2$ corresponds to  low-recoil limit
and in this region the  kaon energy is around a GeV or below. 
Here soft collinear effective theory (SCET) and QCD factorization approaches are not justified properly and become invalid near the zero 
recoil point $q^2 \sim q^2_{max} = \left(m_B - m_{K^*}\right)^2$.  
At low recoil the heavy to light decays can be studied by an operator product expansions  in $1/Q$ where $Q = \left(m_b, \sqrt{q^2}\right)$ 
\textit{i.e.} $\sqrt{q^2}$ is of the order of the mass of the $b$ quark, $m_b$ \cite{dan, feldmann3}. The combination of operator 
product expansion (OPE) with the heavy quark effective theory (HQET) and the use of improved Isgur-Wise form factor relations \cite{dan, dan2} 
allows to obtain the $\bar{B} \rightarrow \bar{K}^* l^+ l^-$ matrix element expansion in the strong coupling and in power corrections suppressed 
by the heavy quark mass, in low recoil.\paragraph*{}

Recently, the observed anomalies associated with $b \to s l ^+ l^-$ processes at LHCb \cite{ lhcb1, lhcb-1a, lhcb-1b, lhcb-1c} have attracted 
a lot of attention to look for new physics both in the context of various new physics models as well as in model-independent ways 
\cite{ matias, ager, huber, bobeth,newref1}.
 In this paper, using scalar leptoquark model, we would like to study the 
   $\bar{B}_d^0 \rightarrow \bar{K}^* l^+ l^-$ processes, which contain quite a large number of clean observables in the full  kinematics except the 
intermediate $q^2$ region. In particular, we are interested to look for the effect of scalar leptoquark on some of the  observables such as 
dilepton mass spectra, lepton-angle distribution and various asymmetries like forward-backward asymmetry and isospin asymmetry. 

The similarities between leptons and quarks lead to the fact that there could exist leptoquarks (LQs), which are
color triplet bosons and carry both lepton ($L$) and baryon ($B$) quantum numbers. Leptoquarks violating both $B$ and $L$ 
numbers are generally considered to be very heavy at the level of ${\cal O}( 10^{15})$ GeV to avoid proton decay. On the other hand
LQs conserving $B$ and $L$ can be light and can have implications in the low energy phenomena.
The existence of leptoquarks has been proposed in many extensions of the SM e.g., Grand Unified Theories (GUTs) \cite{ref7}, Pati-Salam model
\cite{pati}, technicolor models \cite{ref8}, composite scenarios \cite{schrempp}, etc. 
 The spin of leptoquarks could be either one (vector leptoquarks) or zero (scalar leptoquarks).  
 Scalar leptoquarks are encountered in extended technicolor models  and models with compositeness of 
quark and lepton \cite{ref8,schrempp} at TeV scale.  However, in this case the bounds from proton decays may not be relevant 
and leptoquarks may give signatures in other low energy processes \cite{wise}. 
The phenomenology of scalar leptoquark and the contribution to new physics has been quite well studied in 
the literature \cite{wise,davidson, lepto, lq1,newref2}. However, the effect of scalar letoquarks in various observables
associated with $B \to K^* \mu^+ \mu^-$ process is not yet explicitly studied. In Ref. \cite{lepto} model independent constraints on leptoquarks
from $b \to s \l^+ l^-$ processes are obtained. In this paper, we would like to see how the scalar leptoquarks affect these observables and 
whether it would be possible to differentiate between these two scalar LQ models from some of these observables.     
\paragraph*{}

The plan of the paper is as follows. We present a brief discussion on the effective Hamiltonian for $b \rightarrow sl^+ l^-$ processes
in the SM as well as in leptoquark model in Section II. The new physics contributions to
these processes  due to the exchange of scalar 
leptoquarks  and the constraint on leptoquark parameter space from the rare decay mode $B_s \rightarrow \mu^+ \mu^- $ 
have also been discussed. The constraints obtained from $B_s -\bar B_s$ mixing is discussed in Section III.
The  observables associated with the decay modes $\bar{B}_d^0 \rightarrow \bar{K}^* l^+ l^-$ 
are presented in Section IV. Our predicted results  on branching ratio, isospin asymmetry parameter and various form factor
independent observables in the angular distribution are also presented in this section. Section V contains  the summary
and conclusion.  
 
\section{Effective Hamiltonian for  $b \rightarrow sl^+ l^-$ process} 
The effective Hamiltonian  describing the flavour-changing quark level transitions $b \rightarrow sl^+ l^-$ in the standard model 
is given as \cite{ buras}
\bea
{\cal H}_{eff} &=& - \frac{ 4 G_F}{\sqrt 2} V_{tb} V_{ts}^* \Bigg[\sum_{i=1}^6 C_i(\mu) O_i +C_7 \frac{e}{16 \pi^2} \Big(\bar s \sigma_{\mu \nu}
(m_s P_L + m_b P_R ) b\Big) F^{\mu \nu} \nn\\
&&+C_9^{eff} \frac{\alpha}{4 \pi} (\bar s \gamma^\mu P_L b) \bar l \gamma_\mu l + C_{10} \frac{\alpha}{4 \pi} (\bar s \gamma^\mu P_L b)
\bar l \gamma_\mu \gamma_5 l\Bigg]\;,\label{ham}
\eea
where $V_{q q'}$ denote the CKM matrix elements, $G_F$ is the Fermi constant, $\alpha$ is the fine-structure constant, 
$P_{L,R} = (1\mp \gamma_5)/2$ is the chirality projection operator and 
$C_{i}$'s are the Wilson coefficients. The values of the Wilson coefficients  evaluated at the scale $\mu = m_b$
in the  next-to-next-leading order are listed in Table-1. 
\begin{table}[htb]
\begin{center}
\caption{The SM Wilson coefficients evaluated at the scale $\mu = 4.6$ GeV \cite{ kohnda}.}
\begin{tabular}{c  c  c  c  c  c  c  c  c  c }
\hline
\hline
 $C_1$ & $C_2$ & $C_3$ & $C_4$ & $C_5$ & $C_6$ & $C_7^{eff}$ & $C_8^{eff}$ & $C_9$ & $C_{10}$ \\
\hline
 -0.3001  \hspace*{0.22cm}&  1.008 \hspace*{0.22cm}& $-0.0047$ \hspace*{0.22cm}& $-0.0827$ \hspace*{0.22cm}& 0.0003 \hspace*{0.22cm}& 0.0009 
\hspace*{0.22cm}& $-0.2969$ \hspace*{0.22cm}& $-0.1642$ \hspace*{0.22cm}& 4.2607 \hspace*{0.22cm}& $-4.2453$\hspace*{0.22cm} \\
 \hline
 \hline
\end{tabular}
\end{center}
\end{table}

The effective Hamiltonian described above in Eq. (\ref{ham}) will receive additional contributions arising due to the exchange of
leptoquarks.  We will present the modified Hamiltonian  in the presence of scalar leptoquarks in the subsection below. 
\subsection{New physics contribution from scalar leptoquark}
Models with scalar leptoquarks can modify the effective Hamiltonian due to  the exchange of leptoquarks 
and will give measurable deviations from the predictions of the SM in the flavor sector. Here we will consider the minimal 
renormalizable scalar leptoquark model \cite{wise}, containing one single additional representation of $SU(3)\times SU(2)\times U(1)$ 
which does not allow baryon number violation in perturbation theory. There are only two such models which 
 are represented as  $X = (3,2,7/6)$ and $X = (3,2,1/6)$ under the $SU(3)\times SU(2)\times U(1)$ gauge group. 
Here, we are interested to study the effects of  these scalar leptoquarks which potentially contribute to the quark level transition 
$b\rightarrow s l^+ l^-$ and 
constrain the underlying couplings from experimental data on $B_s \rightarrow \mu^+ \mu^-$. Although the details of this
method has been discussed in Refs. \cite{mohanta, mohanta2}, here we will briefly mention about the main points for completeness. 
\paragraph*{}

The interaction Lagrangian for the scalar leptoquark $X = (3,2,7/6)$ couplings to the fermion bilinear  \cite{wise} is
\begin{equation}
\mathcal{L} = -\lambda^{ij}_u \bar{u}_R^i X^T \epsilon L_L^j - \lambda^{ij}_e \bar{e}_R^i X^\dagger Q_L^j + h.c.\;,
\end{equation}
where $i, j$ are the generation indices, $Q_L$  ($L_L$) is the left handed quark (lepton) doublet, $X$ is the scalar leptoquark
doublet, $u_R$ ($e_R$) is the right handed up-type quark (charged lepton) singlet and
$\epsilon = i\sigma_2$ is a $2 \times 2$ matrix.

After expanding the $SU(2)$ indices and performing Fierz transformation, the contribution to the interaction Hamiltonian for 
the process $b\rightarrow s \mu^+ \mu^-$ is
\begin{equation}
\mathcal{H}_{LQ} = \frac{\lambda_\mu^{32} \lambda_\mu^{22 *}}{8M_Y^2} \left[\bar{s}\gamma^\mu (1-\gamma_5)b\right] \left[\bar{\mu}\gamma_\mu (1+\gamma_5)\mu \right] = \frac{\lambda_\mu^{32} \lambda_\mu^{22 *}}{4M_Y^2}\left(\mathcal{O}_9 + \mathcal{O}_{10}\right),
\end{equation}
which can be written  analogous to the  SM effective Hamiltonian as 
\begin{equation}
\mathcal{H}_{LQ} = -\frac{G_f \alpha}{\sqrt{2}\pi}V_{tb}V_{ts}^*\left(C_9^{NP}\mathcal{O}_9 + C_{10}^{NP}\mathcal{O}_{10}\right).
\end{equation}
Thus, one obtains the new Wilson coefficients
\begin{equation}
C_9^{NP} = C_{10}^{NP} = -\frac{\pi}{2\sqrt{2}G_f \alpha V_{tb}V_{ts}^*}\frac{\lambda_\mu^{32} \lambda_\mu^{22 *}}{M_Y^2}.
\end{equation}
Similarly, the corresponding Lagrangian for the coupling of scalar leptoquark $X = (3,2,1/6)$ to the fermion bilinear is
\begin{equation}
\mathcal{L} = -\lambda^{ij}_d \bar{d}_R^i X^T \epsilon L_L^j + h.c., \hspace{2cm}
\end{equation}
Proceeding in the similar manner as done in the previous case, the interaction Lagrangian becomes 
\begin{equation}
\mathcal{H}_{LQ} = \frac{\lambda_s^{22} \lambda_b^{32 *}}{8M_V^2} \left[\bar{s}\gamma^\mu (1+\gamma_5)b\right] \left[\bar{\mu}\gamma_\mu 
(1-\gamma_5)\mu \right] = \frac{\lambda_s^{22} \lambda_b^{32 *}}{4M_V^2}\left(\mathcal{O}_9^\prime - \mathcal{O}_{10}^\prime \right),
\end{equation}
where $\mathcal{O}_9^\prime$ and $\mathcal{O}_{10}^\prime$ are dimension-six operators obtained
from ${\cal O}_9$ and ${\cal O}_{10}$ by the replacement $P_L \leftrightarrow P_R$ 
and their respective new Wilson coefficients due to the exchange of the leptoquark $X = (3,2,1/6)$ are given as
\begin{equation}
C_9^{\prime NP} = -C_{10}^{\prime NP} = \frac{\pi}{2\sqrt{2}G_f \alpha V_{tb}V_{ts}^*}\frac{\lambda_s^{22} \lambda_b^{32 *}}{M_V^2}.
\end{equation}

After having the new Wilson coefficients in hand,  we now  proceed to constrain the combination of LQ couplings by comparing the  
theoretical \cite{gorbahn} and  experimental  branching ratios \cite{lhcb2, lhcb3, lhcb4} of $B_s \to \mu^+ \mu^-$, as these new 
coefficients contribute to the  $B_s \to \mu^+ \mu^-$ process as well.
Furthermore, we  require that  each individual leptoquark contribution  to the branching ratio does not exceed the 
experimental result. The constraint on leptoquark parameter space has been extracted in \cite{mohanta, mohanta2}, therefore here we will 
simply quote the results.

The allowed region
in $r-\phi^{NP}$ plane which is compatible with the $1\sigma$ range of
the experimental data is 
$0\leq r \leq 0.1 $ for the entire range of $\phi^{NP}$, i.e.,
\bea
 0\leq r \leq 0.1\;, ~~~~{\rm for}~~~~0 \leq \phi^{NP} \leq 2 \pi \;,
 \eea
where $r$ and $\phi^{NP}$ are defined as
\be
r e^{i \phi^{NP}}= (C_{10}^{NP}-C_{10}^{'NP})/C_{10}^{SM}\;.\label{bound}
\ee
However, in our analysis we will use relatively mild constraint, consistent with both measurement of 
${\rm BR}(B_s \to \mu^+ \mu^-)$ and ${\rm BR}(\bar B_d^0 \to X_s \mu^+ \mu^-)$ \cite{mohanta} as
 \bea
 0\leq r \leq 0.35\;, ~~~~{\rm with}~~~~\pi/2 \leq \phi^{NP} \leq 3 \pi/2\;.\label{r-bound1}
 \eea
 It should be noted that the use of this limited range of CP phase, i.e.,  ($\pi/2 \leq \phi^{NP} \leq 3 \pi/2$) 
is an assumption to have a relatively larger value of $r$. The constraint on $r$   can be translated to obtain the 
bounds for the leptoquark couplings using Eqs. (5), (8) and (11) as
 \bea
 0 \leq \frac{|\lambda^{32} {\lambda^{22}}^*|}{M_S^2} \leq 5 \times 10^{-9} ~ {\rm GeV}^{-2}~~~~{\rm for}~~~~\pi/2 \leq \phi^{NP} \leq 3 \pi/2\;.\label{scale1}
 \eea
\section{Bound from $B_s-{\bar B}_s$ mixing}
Now we will obtain the constraint on the leptoquark couplings from the mass difference  between the $B_s$ meson mass eigenstates ($\Delta M_s$),
which characterizes the $B_s - \bar B_s$ mixing phenomena. In the SM, $B_s-\bar B_s$ mixing proceeds to an excellent approximation
through the box diagram with internal top quark and $W$ boson exchange, and
 the effective Hamiltonian describing the $\Delta B=2$
transition  is given by \cite{lim} 
\be {\cal H}_{eff}=\frac{G_F^2}{16 \pi^2}~ \lambda_t^2~ M_W^2 S_0(x_t)\eta_B
(\bar s b)_{V-A}(\bar s b)_{V-A}\;, \ee where 
$\lambda_t=V_{tb} V_{ts}^*$, $\eta_B$ is the QCD correction factor and $S_0(x_t)$ is
the loop function \be S_0(x_t)=\frac{4 x_t -11 x_t^2
+x_t^3}{4(1-x_t)^2} - \frac{3}{2} \frac{\log x_t x_t^3}{(1-x_t)^3}\;,
\ee with $x_t=m_t^2/M_W^2$. Thus, the $B_s - \bar B_s$ mixing
amplitude in the SM can be written as \be M_{12}^{SM}=\frac{1}{2
M_{B_s}} \langle \bar B_s|{\cal H}_{eff}| B_s \rangle = \frac{G_F^2}{12
\pi^2} M_W^2~ \lambda_t^2~ \eta_B~ \hat B_s f_{B_s}^2 M_{B_s} S_0(x_t)\;,
\label{sm} \ee 
where the vacuum insertion method has been used to
evaluate the matrix element \be\langle \bar B_s|(\bar s
\gamma^\mu (1-\gamma_5) b)(\bar s \gamma_\mu (1-\gamma_5) b)|B_s \rangle = \frac{8}{3} \hat B_s f_{B_s}^2
M_{B_s}^2\;.\label{mix}\ee The corresponding mass difference is
related to the mixing amplitude through $\Delta M_s = 2 |M_{12}|$.
Now using the particle masses from \cite{pdg}, $\eta_B = 0.551$, the Bag parameter $\hat B_{B_s}=1.320 \pm 0.017 \pm 0.030$ 
the decay constant $f_{B_s}=225.6 \pm 1.1 \pm 5.4 $,   t-quark mass $m_t =165.95$ from \cite{ckmfitter}, 
we obtain the value of
$\Delta M_s$ in the SM as
\bea
\Delta M_s^{SM} = (17.426\pm 1.057)~ {\rm ps^{-1}},
\eea
which is in  good agreement with the experimental result \cite{pdg}
 \bea
\Delta M_s = 17.761 \pm 0.022~ {\rm ps^{-1}}.\label{mass-diff}
\eea
However, the central value of the theoretical 
prediction deviates from the corresponding experimental value. The ratio of these
two results yields
\bea 
\Delta M_s/\Delta M_s^{SM}=1.019 \pm 0.062\;,\label{ratio}
\eea
which is consistent with one, but it does not completely rule out 
the possibility of new physics in $B_s - \bar {B}_s$ mixing. 
\begin{figure}[htb]
\includegraphics[width=15.0cm,height=4.5 cm]{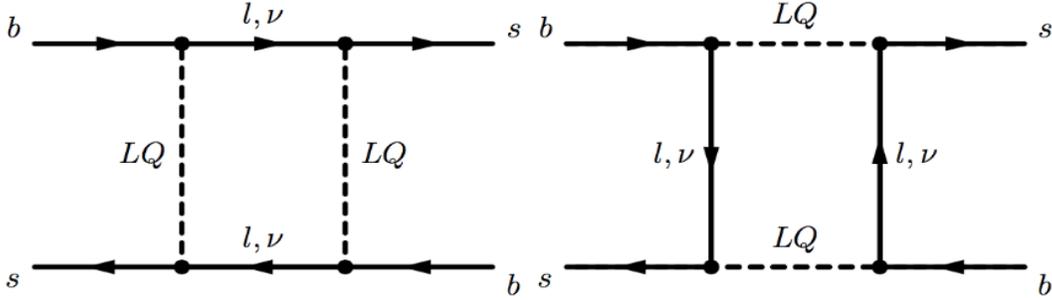}
\caption{Box diagram for $B_s-\bar B_s$  mixing phenomenon in the leptoquark model.}
\end{figure}
The mixing amplitude  receives additional contribution due to the flow of leptoquark and charged lepton/neutrino in the 
box diagram as shown in Fig.1. For $X(3,2,7/6)$ LQ, there will be contribution coming only from charged lepton in the loop whereas for
 $X(3,2,1/6)$ both charged lepton and neutrino will contribute to the mixing amplitude.  

The effective Hamiltonian due to the leptoquark $X(3,2,7/6)$ and charged lepton in the loop is given by
\be
{\cal H}_{eff}=\sum_{i=e,\mu,\tau} \frac{(\lambda^{bi} {\lambda^{si}}^{*})^{2}}{128 \pi^2}\frac{1}{ M_{S}^2}~I
\left (\frac{m_i^2}{M_S^2} \right )(\bar b \gamma^\mu P_L s) (\bar b \gamma_\mu P_L s)\;,
\ee 
where the loop function  $I(x)$ is given as 
\be
I(x)=\frac{1-x^2+2x \log x}{(1-x)^2},
\ee
which is always very close to $I(0)=1$.
For $X(3,2,1/6)$ contribution there will be charged lepton as well as neutrinos in the loop and the corresponding effective Hamiltonian
becomes 
\be
{\cal H}_{eff}=\sum_{i=e,\mu,\tau} \frac{({\lambda^{bi}}^{*} \lambda^{si})^{2}}{128 \pi^2}\left [ \frac{1}{ M_{S}^2}~I
\left (\frac{m_i^2}{M_S^2} \right )+\frac{1}{ M_S^2} \right ] (\bar b \gamma^\mu P_R s) (\bar b \gamma_\mu P_R s)\;.
\ee 
\begin{figure}[htb]
\centering
\includegraphics[width=7.5cm,height=5.5cm]{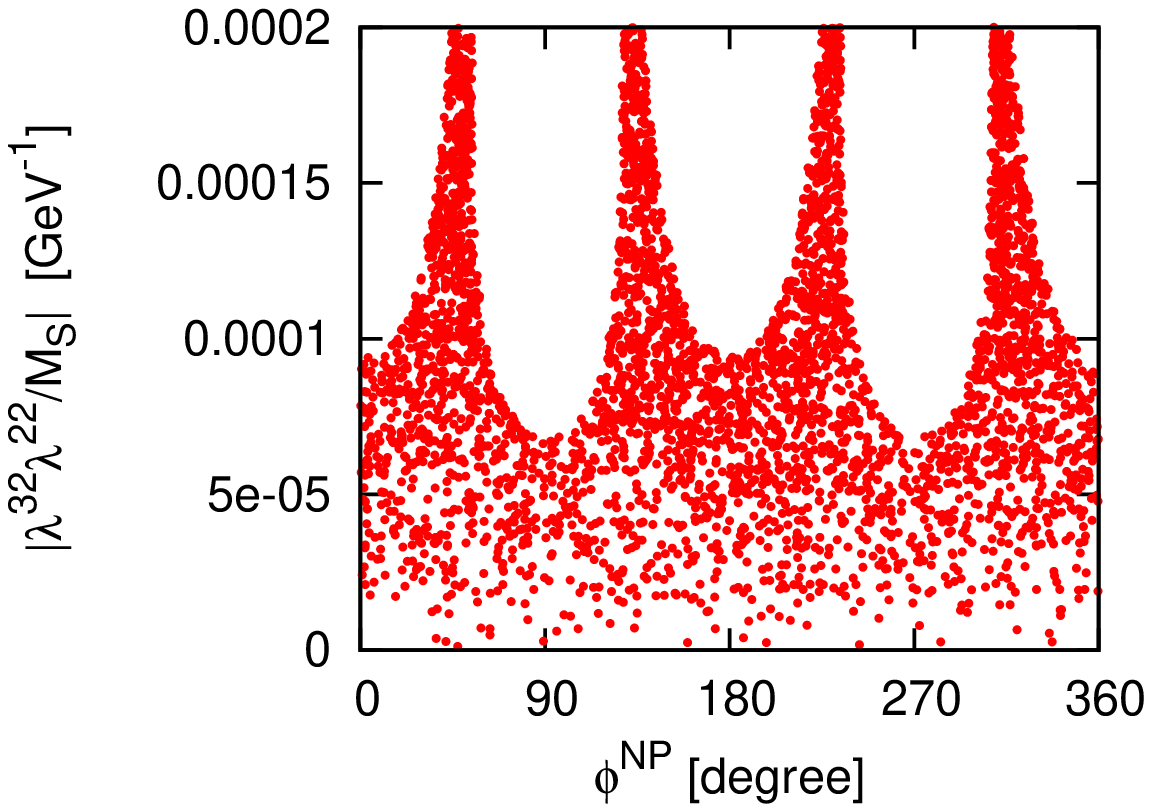}
\quad
\includegraphics[width=7.5cm,height=5.5cm]{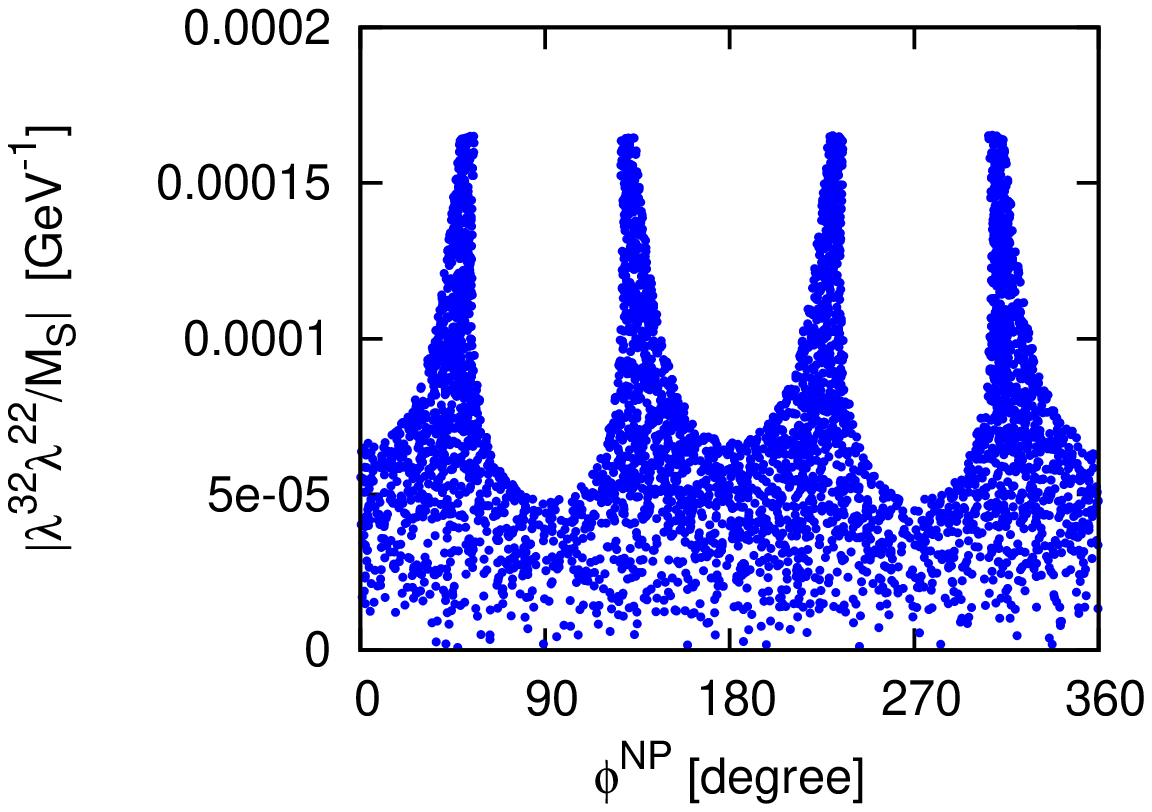}
\caption{The allowed parameter space for the leptoquark couplings  in the $|\lambda^{32} \lambda^{22}/M_S|$  vs. $\phi^{NP}$ plane
obtained from the mass difference between $B_s$ meson mass eigenstates ($\Delta M_s$). The left panel corresponds to bounds on $X(3,2,7/6)$ and right panel is for
$X(3,2,1/6)$ couplings.}
\end{figure} 

To obtain the constraints on the leptoquark coupling, we require that individual leptoquark  contribution to the
mass difference  does not exceed the $1\sigma$ range of the experimental value. Since we are interested to obtain the bounds on 
$\lambda^{b\mu }$ and $\lambda^{s\mu }$ couplings, we consider the muon contribution to
the mixing amplitude.  Neglecting the mass of muon and using Eq. (\ref{mix}), we obtain the contribution  due to leptoquark exchange as
\bea
M_{12}^{LQ} &= & \frac{{(\lambda^{32 }}^{*} {\lambda^{2 2}})^2}{192 \pi^2 M_S^2} \eta_B \hat B_{B_s} f_{B_s}^2 M_{B_s}\;,~~~~~~{\rm for}~X(3,2,1/6)\nn\\
M_{12}^{LQ} &= & \frac{(\lambda^{32} {\lambda^{2 2}}^*)^2}{384 \pi^2 M_S^2} \eta_B \hat B_{B_s} f_{B_s}^2 M_{B_s}\;,~~~~~~{\rm for}~X(3,2,7/6).
\eea
Thus, including both SM and leptoquark couplings the total contribution to mass difference is 
given as
\be
\Delta M_s = \Delta M_s^{SM} \left | \left [1 + \frac{c}{16 G_F^2 V_{tb}^2 V_{ts}^{*2} m_W^2 S_0(x_t)}
 \left (\frac{(\lambda^{32} {\lambda^{22}}^{*})^2}{M_S^2}  \right )\right ]\right |\;,
\ee 
where the constant $c= 1 $ for $X(3,2,1/6)$ and $1/2$ for  $X(3,2,7/6)$.  Now varying the ratio of  mass difference $(\Delta M_s/\Delta M_s^{SM}) $
   within its 
$1 \sigma $ allowed range  (\ref{ratio}),  we obtain the constraint on $|\lambda^{32} \lambda^{22}/M_S|$ as shown in 
 Fig. 2,  where the left plot corresponds to constraint on $X(3,2,7/6)$ and the right plot shows the constraint on $X(3,2,1/6)$ couplings.
From the figure, the bounds on $|\lambda^{32} \lambda^{22}/M_S|$   for the entire allowed range of $\phi^{NP}$ are found to be 
\bea
0 \leq \left |\frac{\lambda^{32} \lambda^{22}}{M_S} \right |\leq 7.5 \times 10^{-5}~{\rm GeV}^{-1}\;,~~~~~~~~ {\rm for}~X(3,2,7/6),\nn\\
0 \leq \left |\frac{\lambda^{32} \lambda^{22}}{M_S} \right | \leq 5.0 \times 10^{-5}~{\rm GeV}^{-1}\;,~~~~~~~~~ {\rm for}~X(3,2,1/6).\label{scale}
\eea 
It should be noted that using the $B_s -\bar B_s$ mass difference, we obtained the bounds  on $|\lambda^{32} \lambda^{22}|/M_S$,
whereas using the $B_s \to \mu \mu $ and $B_d \to X_s \mu \mu$ data the bounds on  $|\lambda^{32} \lambda^{22}|/M_S^2$  have been obtained.
So to correlate these two results, we need to know the mass of the scalar leptoquark $M_S$. Recently CMS collaboration \cite{cms-lq1} 
with 8 TeV data set excluded the first generation leptoquarks with masses less than 1010 (850) GeV for $ \beta= 1.0 ~(0.5)$, where
$\beta$ is the branching fraction of a leptoquark decaying to a charged lepton and a quark. The second generation scalar LQs are excluded
with masses less than 1080 (760) GeV for $ \beta= 1.0 ~(0.5)$. They also ruled out at $95\%$ confidence level the single production of first
generation LQs with coupling and branching fraction of 1.0 for masses below 1730 GeV and for second generation for masses below
530 GeV \cite{cms-lq2}. ATLAS Collaboration excluded at $95\%$ C.L. the scalar leptoquarks with masses upto 1050 GeV for first generation LQs (i.e.,
$m_{LQ1} < 1050$ GeV), $m_{LQ2} < 1000$ GeV for second generation LQs and $m_{LQ3} < 625$ GeV for third generation LQs \cite{atlas-lq}.
Hence, if we  scale the couplings obtained from $B_s - \bar B_s$ mass difference  for a benchmark  leptoquark mass of 1 TeV, the bounds in Eq. 
(\ref{scale})  can be translated  as 
\bea
0 \leq \left |\frac{\lambda^{32} \lambda^{22}}{M_S^2}\right | 
 \leq 7.5 \times 10^{-8}~{\rm GeV}^{-2},~~~~~~~~ {\rm for}~X(3,2,7/6)\nn\\
0 \leq \left |\frac{\lambda^{32} \lambda^{22}}{M_S^2}\right | 
\leq 5.0 \times 10^{-8}~{\rm GeV}^{-2},~~~~~~~~ {\rm for}~X(3,2,1/6).
 \eea
Since these bounds are reasonably higher than those of  obtained from $B_s \to \mu \mu$ and $B_d \to X_s \mu \mu$ , 
we will use the bounds (\ref{scale1}) as mentioned in the previous section, in our analysis. 
\section{Analysis of $B_d^0 \rightarrow K^* l^+ l^-$ processes}  

Here we will consider the decay modes $B_d^0 \to K^* l^+ l^-$. At the quark level, these processes
proceed through the FCNC transition $b \to sl^+ l^- $, which occurs only through loops in the
SM, and hence they are quite suitable to look for new physics. Moreover,
the dileptons present in these processes allow one to formulate several useful observables which can serve
as a testing ground to decipher the presence of new physics.

The transition amplitude for these processes can be obtained using the effective Hamiltonian presented in Eq. (\ref{ham}). 
The matrix elements of the various hadronic currents between the
initial $B$ meson and the final $K^*$ vector meson  can be parameterized in terms of  seven form factors 
by means of a narrow width approximation. The relevant form factors \cite{ball2} are given as
 \beqa
   \langle K^* \left(p_{K^*}\right)|\bar{s} \gamma _\mu  P_{L,R}  b | B\left(p\right)\rangle = i\epsilon_{\mu \nu \alpha \beta} \epsilon^{\nu *} p^\alpha q^\beta 
\frac{V(s)}{m_B + m_{K^*}} \mp \frac{1}{2} \Bigg( \epsilon^*_\mu (m_B + m_{K^*}) A_1(s) \nn\\ 
-(\epsilon^* \cdotp q)(2p-q)_\mu \frac{A_2(s)}{m_B + m_{K^*}} - \frac{2m_{K^*}}{s} (\epsilon^* \cdotp q) \left[ A_3(s) - A_0(s)\right] q_\mu \Bigg),\nn\\
 \langle K^* \left(p_{K^*}\right)| \bar{s}  i \sigma_{\mu \nu}q^\nu P_{L,R}b | B\left(p\right)\rangle =  -i\epsilon_{\mu \nu \alpha \beta} 
\epsilon^{\nu *} p^\alpha  q^\beta \emph{T}_1(s) \pm \frac{1}{2} \Bigg( \Big[ \epsilon^*_\mu (m^2_B - m^2_{K^*})\nn\\ 
 - (\epsilon^* \cdotp q)(2p-q)_\mu \Big]  \emph{T}_2(s)    + (\epsilon^* \cdotp q) \Bigg[ q_\mu -\frac{s}{(m^2_B - m^2_{K^*})} (2p-q)_\mu \Bigg] 
\emph{T}_3(s)\Bigg),
\eeqa
 where $q = p_{l^+} + p_{l^-}$, $s = q^2$ and $\epsilon^\mu$ is the polarization vector of $K^*$. The form factors $A_1$, $A_2$ and $A_3$
are related to each other through
\begin{equation}
 A_3(s) = \frac{(m_B + m_{K^*})}{2m_{K^*}} A_1(s) - \frac{(m_B - m_{K^*})}{2m_{K^*}} A_2(s).
\end{equation}  
The amplitude for the process $B \to K^*(\to K \pi) l^+ l^-$ can be represented by seven transversity amplitudes,
$A_{\perp,\parallel,0}^{L,R}$ and $A_t$. The explicit form of these amplitudes (up to corrections $O(\alpha_s)$ are
presented in Appendix A (B) for low $q^2$ (high $q^2$) region.

Assuming the $\bar{K}^{* 0}\rightarrow K^-\pi^+$ to be on the mass shell, the full angular distribution of the decay 
$\bar{B} \rightarrow \bar{K}^{* 0}\left(\rightarrow K^-\pi^+\right) l^+ l^-$ can be described by four independent kinematic variables, 
the lepton-pair invariant mass  and the three angles $\theta_{K^*}, \theta_l$ and $  \phi$. 
The differential decay distribution in terms of three variables can be written as \cite{hiller2, egede,egede2}
 \begin{equation}
 \frac{d^4\Gamma}{dq^2~ d\cos\theta_l ~d\cos\theta_{K^*}~ d\phi} = \frac{9}{32\pi} J\left(q^2, \theta_l, \theta_{K^*}, \phi\right)\;,
 \end{equation}
 where the lepton spins have been summed over.
 Here $q^2$ is the dilepton invariant mass squared, $\theta_l$ is defined as the angle between the negatively charged lepton and 
the $\bar{B}$ in the dilepton frame, $\theta_{K^*}$ is the angle between $K^-$ and $\bar{B}$ in the $K^-\pi^+$ center 
of mass system and $\phi$ is given by the angle between the normals of the $K^-\pi^+$ and the dilepton $(l^+l^-)$ planes.
 The full kinematically physical region phase space is given by 
 \begin{equation}
 4m^2_l \leqslant q^2 \leqslant \left(m_B - m_{K^*}\right)^2,\hspace{0.6cm} -1\leqslant \cos\theta_l \leqslant 1,\hspace{0.6 cm} -1
\leqslant \cos\theta_{K^*} \leqslant 1,\hspace{0.7cm} 0\leqslant \phi \leqslant 2\pi.
 \end{equation}
 where $m_B$, $m_{K^*}$, $m_l$ are the masses of $B$ meson, $K^*$ and lepton respectively.
More explicitly the dependence of the decay distribution on the three angles can be written as
\beqa
 J\left(q^2, \theta_l, \theta_{K^*}, \phi\right) &= & J^s_1 \sin^2\theta_{K^*} + J^c_1 \cos^2\theta_{K^*} + \left(J^s_2 \sin^2\theta_{K^*} 
+ J^c_2 \cos^2\theta_{K^*}\right) \cos2\theta_l \nn\\
& +& J_3 \sin^2\theta_{K^*} \sin^2\theta_l \cos2\phi + J_4 \sin2\theta_{K^*} \sin2\theta_l \cos\phi 
  +  J_5 \sin2\theta_{K^*} \sin\theta_l \cos\phi\nn\\
& +&(J_6^s \sin^2\theta_{K^*} +J_6^c \cos^2\theta_{K^*})\cos\theta_l
 + J_7 \sin2\theta_{K^*} \sin\theta_l \sin\phi 
\nn\\ 
& +&  J_8 \sin2\theta_{K^*} \sin2\theta_l \sin\phi + J_9 \sin^2\theta_{K^*} \sin^2\theta_l \sin2\phi\;,
\eeqa
 where the coefficients $J_i^{(a)} = J_i^{(a)}\left(q^2\right)$ for $i = 1,....,9$ and $a = s,c$ are functions of the dilepton mass,
 and are expressed in terms of the transversity amplitudes $A_0$, $A_\parallel$, $A_\perp$, and $A_t$ as given in Appendix C.

The dilepton invariant mass spectrum for $\bar{B} \rightarrow \bar{K}^* l^+ l^-$ decay after integration over all angles \cite{hiller2} is given by
 \begin{equation}
 \frac{d\Gamma}{dq^2} = \frac{3}{4} \left(J_1 - \frac{J_2}{3}\right).
 \end{equation}
where $J_i = 2J_i^s + J_i^c$. 
An interesting observable to look for new physics is the zero crossing of forward-backward asymmetry $A_{FB}$, which can be obtained after 
integrating the 4-differential distribution over $\phi$ and $\theta_{K^*}$ angles and is defined as \cite{hiller2}
 \beqa
  A_{FB}\left(q^2\right) & = & \left[ \int_{-1}^0 d\cos\theta_l \frac{d^2\Gamma}{dq^2 d\cos\theta_l}
 - \int_{0}^1 d\cos\theta_l \frac{d^2\Gamma}{dq^2 d\cos\theta_l}\right] \Big{/}
  \frac{d\Gamma}{d q^2}\nn \\&  =& -\frac{3}{8} \frac{J_6}{d\Gamma/dq^2}\;.
   \eeqa
The longitudinal and transverse polarization fraction of the $K^*$ meson can be defined in terms of the transversity amplitudes as
\bea
F_L(s)= \frac{|A_0|^2}{|A_0|^2+|A_\parallel|^2+|A_\perp|^2}\;,~~~~~~~F_T(s)= \frac{|A_\parallel|^2+|A_\perp|^2}{|A_0|^2+|A_\parallel|^2+|A_\perp|^2}\;,
\eea
and in terms of the angular coefficients $J_i$'s these observables can be expressed as 
  \cite{dyk}
 \begin{equation}
 F_L\left(s\right) = \frac{3J_1^c-J_2^c}{4d\Gamma/dq^2}\;, \hspace{1.5 cm}
 F_T\left(s\right) = 1-F_L(s)\,, 
 \end{equation}
 so that one can define the ratio of $K^*$ polarization fraction $\alpha_{K^*}$ as \cite{matias2}
 \begin{equation}
 \alpha_{K^*}\left(s\right) = \frac{2F_L}{F_T} - 1 = \frac{-J_2}{2J_2^s}\;.\hspace{4cm}
 \end{equation} 
 The transverse asymmetries are given as \cite{matias2}
 \beqa
  A_T^{(1)}(s) &=& \frac{-2 {\rm Re}\left(A_{\parallel}A^*_\perp\right)}{|A_\perp|^2 + |A_\parallel|^2},\nn\\
  A_T^{(2)}\left(s\right) & = & \frac{J_3}{2J^s_2},\nn\\
  A_T^{(3)}\left(s\right) & = &\left(\frac{4\left(J_4\right)^2 + \beta^2_l\left(J_7\right)^2}{-2J^c_2\left(2J^s_2+J_3\right)}\right)^{1/2},\nn\\
 A_T^{(4)}\left(s\right) & =& \left(\frac{4\left(J_8\right)^2 + \beta^2_l\left(J_5\right)^2}{4\left(J_4\right)^2 + \beta^2_l\left(J_7\right)^2}\right)^{1/2},
\nn\\
 A_T^{(5)}\left(s\right) & = & \frac{|A^L_\perp A_\parallel ^{R *}+A^L_\parallel A_\perp ^{R *}|}{|A_\perp |^2 + |A_\parallel |^2}, \nn\\
  A_{\rm Im}\left(s\right) &=& \frac{J_9}{d\Gamma/dq^2}.
 \eeqa
Another set of interesting observables are the six form factor independent (FFI) observables  \cite{matias4}, which are given by 
 \beqa
 P_1\left(s\right) &=& \frac{J_3}{2J_2^s}\;, \hspace{1.5cm}  P_2\left(s\right)  =  \beta_l\frac{J^s_6}{8J_2^s}\;,\hspace{1.5cm}
 P_3\left(s\right)  = - \frac{J_9}{4J_2^s}\;,\nn\\
 P_4\left(s\right) & = & \frac{\sqrt{2}J_4}{\sqrt{-J_2^c\left(2J_2^s-J_3\right)}}\;,\hspace{1cm}
 P_5\left(s\right) =  \frac{\beta_l J_5}{\sqrt{-2J_2^c\left(2J_2^s+J_3\right)}}\;,\nn\\
 P_6\left(s\right) & = & - \frac{\beta_l J_7}{\sqrt{-2J_2^c\left(2J_2^s-J_3\right)}}\;.
 \eeqa
A slightly modified set of clean observables $P_{4,5,6}'$ which are related to $P_{4,5,6}$ through the relations \cite{matias}
\bea
P_4' \equiv P_4 \sqrt{1-P_1} = \frac{J_4}{\sqrt{-J_2^c J_2^s}}\;,\nn\\
P_5' \equiv P_5 \sqrt{1+P_1} = \frac{J_5}{2\sqrt{-J_2^c J_2^s}}\;,\nn\\
P_6' \equiv P_6 \sqrt{1-P_1} = \frac{-J_7}{2\sqrt{-J_2^c J_2^s}}\;.
\eea
 \subsection{Observables in the large recoil}
After getting familiar with the different observables, we now proceed to study these observables  in the large recoil limit.
For that we need to know the associated form factors for  $B \to K^* l^+ l^- $ process. 
In the heavy quark limit the QCDf form factors obey symmetry relations and at leading order in the $1/E$ expansion, 
they can be expressed in terms of two universal soft non-perturbative form factors $\xi_\perp$ and $\xi_\parallel$. 
In order to calculate the universal form factors we use the QCDf  scheme  \cite{hiller2, feldmann2}, where they are
expressed as
 \begin{equation}
 \xi_\perp(E_{K^*}) = \frac{m_B}{m_B + m_{K^*}} V(q^2)\;, \hspace{0.5cm} \xi_\parallel(E_{K^*}) = \frac{m_B + m_{K^*}}{2E_{K^*}}A_1 (q^2)
- \frac{m_B - m_{K^*}}{m_B}A_2(q^2)\;.
 \end{equation}
 The $q^2$ dependence of the  form factors $V,~A_1, ~A_2$  can be parameterized as 
 \begin{eqnarray}
 V\left(q^2\right))&=& \frac{r_1}{1-q^2/m^2_R} + \frac{r_2}{1-q^2/m^2_{fit}}\;,\nn\\
  A_1\left(q^2\right) &=&\frac{r_2}{1-q^2/m^2_{fit}}\;,\nn\\
 A_2\left(q^2\right)& =& \frac{r_1}{1-q^2/m^2_{fit}} + \frac{r_2}{\left(1-q^2/m^2_{fit}\right)^2}\;.
 \end{eqnarray}
The values of the parameters involved in the calculation of form factors are taken from \cite{ball3}. 
The $T_i$ formfactors are related to the universal form factors $\xi_{\perp, \parallel}$ as
\bea
T_1(q^2)=\xi_{\perp}(E_{K^*})\;,~~~~T_2(q^2)=\frac{2E_{K^*}}{m_B}\xi_{\perp}(E_{K^*})\;,~~~~T_3(q^2)=\xi_{\perp}(E_{K^*})-\xi_{\parallel}(E_{K^*})\;.
\eea
 \begin{table}[htb]
\caption{Summary of the values of various input parameters.}
\begin{center}
\begin{tabular}{c c c c c}
\hline
\hline
& $\alpha$ \hspace{1cm} &1/137 \hspace{2cm} & $a_{1,\perp}$ \hspace{1cm} &0.10\\
& $\alpha_s(M_z)$ \hspace{1cm} & 0.1184 \hspace{2cm} & $a_{2,\perp}$ \hspace{1cm} &0.13\\
& $m_{b, ps}$ \hspace{1cm} & 4.6 GeV \hspace{2cm} & $a_{1, \parallel}$ \hspace{1cm} &0.10\\
& $m_{c, pole}$ \hspace{1cm} & 1.4 GeV \hspace{2cm} & $a_{2, \parallel}$ \hspace{0.8cm} &0.09\\
& $f_{{K^*},\perp}$ \hspace{0.8cm} & 0.185 GeV \hspace{2cm} &$\lambda$\hspace{1cm} & 0.22537$\pm$0.0006\\
& $f_{{K^*},\parallel}$ \hspace{1cm} & 0.220 GeV \hspace{2cm} &$A$\hspace{1cm} & $0.814^{+0.023}_{-0.024}$\\
& $f_B$ \hspace{1cm} &  0.2 GeV  \hspace{2cm} &$\bar{\rho}$\hspace{1cm} & 0.117 $\pm$ 0.021 \\
&  &   \hspace{2cm} &$\bar{\eta}$\hspace{1cm} & 0.353 $\pm$ 0.013 \\
 \hline
 \hline
\end{tabular}
\end{center}
\end{table}

After getting familiar with the different observables and the associated form factors for  $B \to K^* l^+ l^- $ 
processes in the high recoil limit, we now proceed for numerical estimation.
The masses of particles and  the lifetime of $B$ meson  are taken from \cite{pdg}. For the leptoquark
couplings we use a representative value for the parameter $r$ as $r=0.3$ and vary the associated phase between $\pi/2 \leq \phi^{NP}
\leq 3 \pi/2$. Furthermore, we will present most of the the results only for $X(3,2,7/6)$ LQ and only a few representative plots
for $X(3,2,1/6)$.
The values of quark masses and all the input parameters used in our analysis are listed in Table-II.

 \begin{figure}[htb]
\includegraphics[width=7.5cm,height=5.5cm]{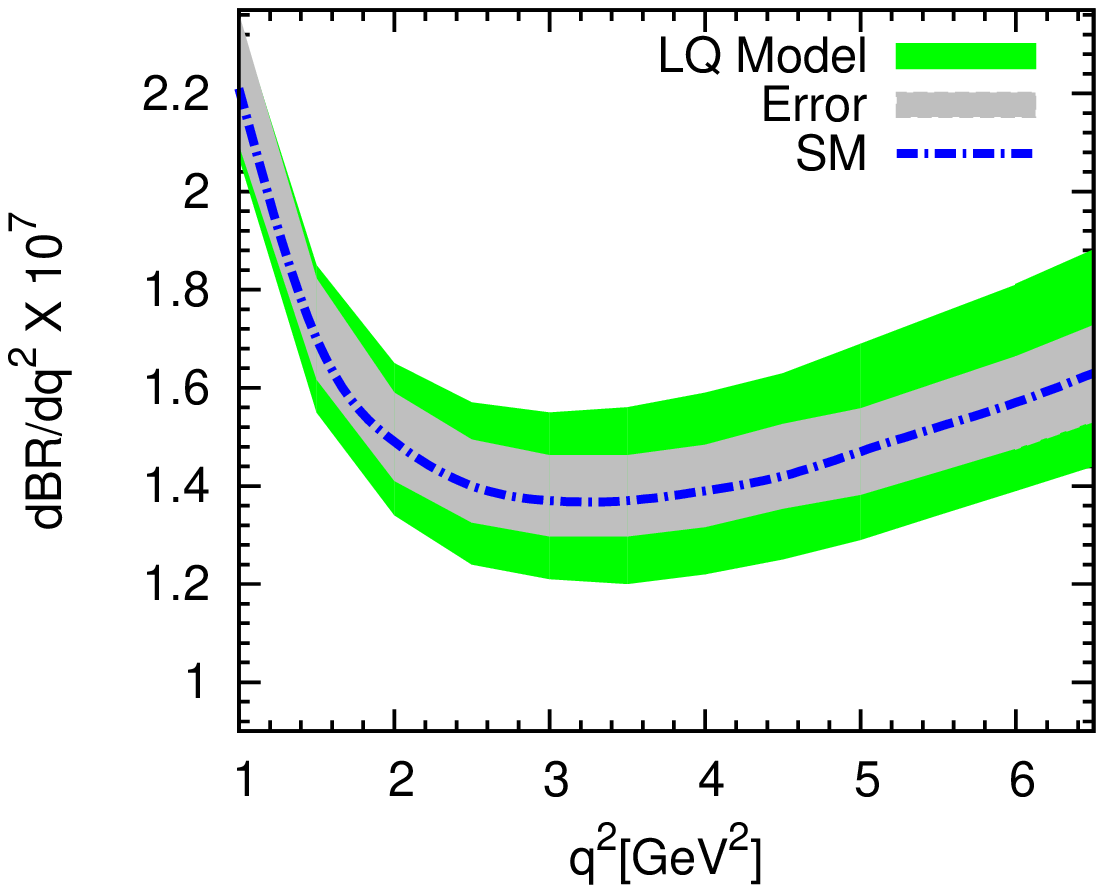}
\quad
\includegraphics[width=7.5cm,height=5.5cm]{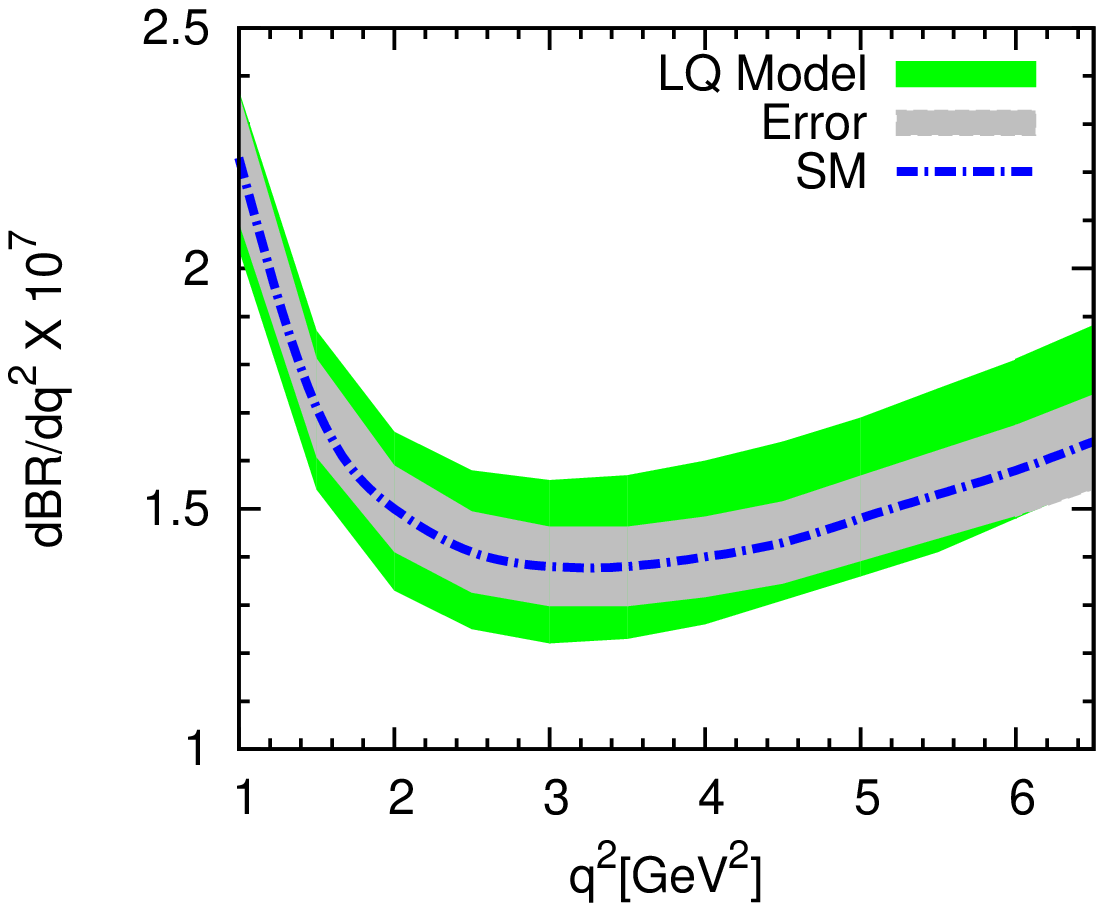}
\includegraphics[width=7.5cm,height=5.5cm]{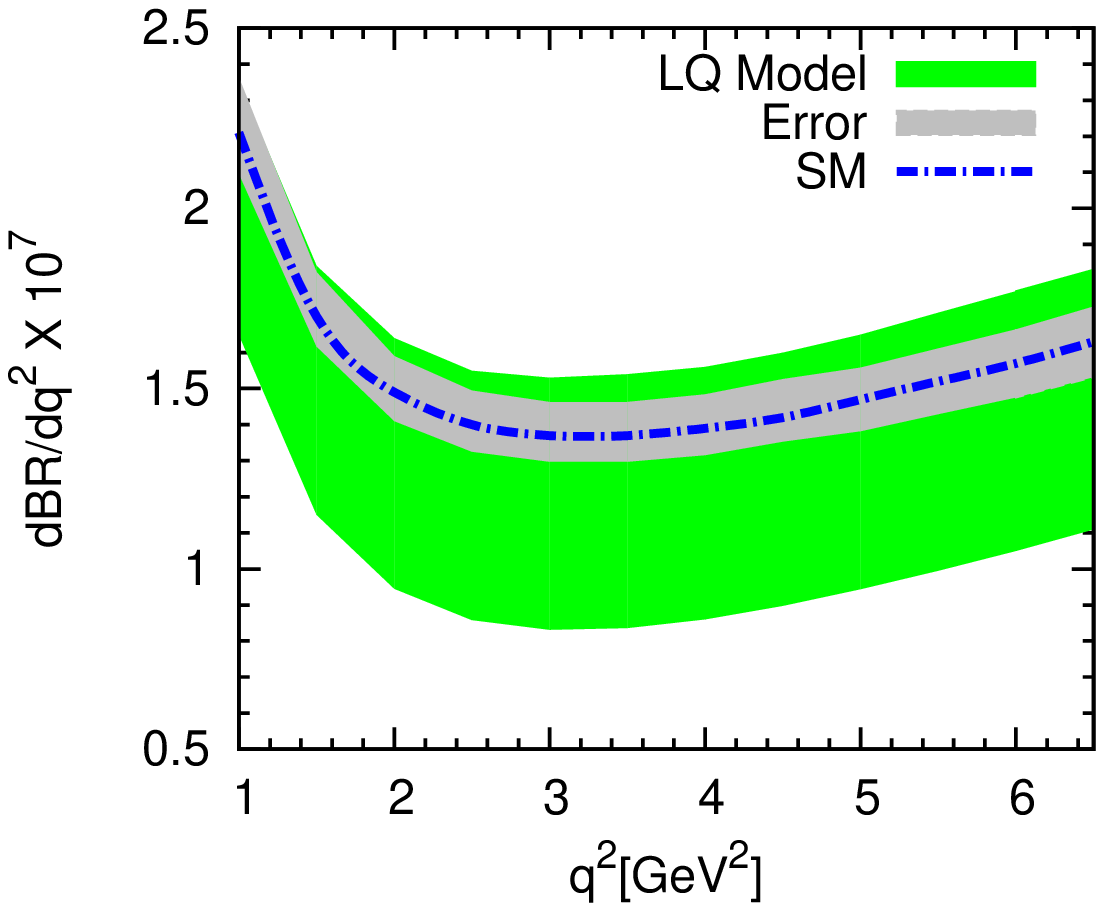}
\quad
\includegraphics[width=7.5cm,height=5.5cm]{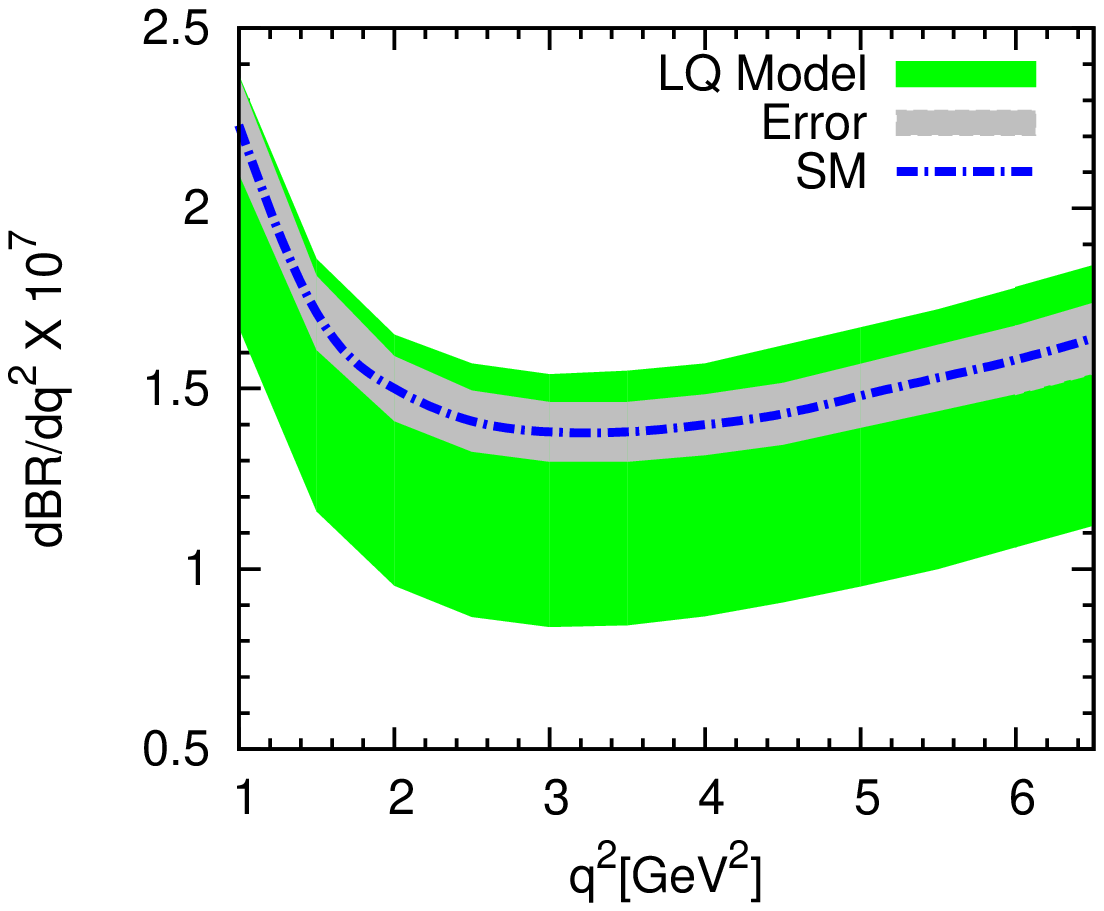}
\caption{The variation of branching ratios of $\bar{B} \rightarrow \bar{K}^* \mu^+ \mu^-$ (left panel) and  $\bar{B} \rightarrow \bar{K}^* e^+ e^-$ 
(right panel) with low $q^2$ in the standard model and in leptoquark mode. The green bands are due to the leptoquark contributions and  
the gray bands represent the theoretical uncertainties from the input parameters in the SM. The plots in the top panel are 
for $X(3,2,7/6)$ and those in the bottom panel are for $X(3,2,1/6)$.}
\end{figure} 
In Fig. 3, we show the variation of the branching ratios for $\bar{B} \rightarrow \bar{K}^* \mu^+ \mu^-$ (left panel) 
and $\bar{B} \rightarrow \bar{K}^* e^+ e^-$ (right panel) in the low $q^2$  region.  The plots in the top panel are 
for $X(3,2,7/6)$ and those in the bottom panel are for $X(3,2,1/6)$. The variation of the longitudinal and transverse polarization 
fractions of $K^*$ has been shown in Fig. 4 and that of forward-backward asymmetry in   
Fig. 5. 
From these figures one can see that the affect of the LQs $X(3,2,7/6)$ and $X(3,2,1/6)$ are quite
different and one can easily differentiate between these two models from the measured values of the $K^*$ polarization
fractions $F_L(q^2)$ and $F_T(q^2)$. 
The transverse asymmetry parameters $A_T^{(3)}$, $A_T^{(4)}$,  $A_{\rm Im}$  and $K^*$ polarization factor $\alpha_{K^*}$ variations with $q^2$ are
presented in Fig. 6.  The variation of form factor independent observables as a function of dimuon invariant mass squared have shown in 
Fig. 7.  The total branching ratios and the asymmetries integrated over the range $q^2 \in [1, 6]~{\rm GeV^2}$ are 
presented in Table III and the allowed range of transverse asymmetry and the form factor independent observables are given in Table IV.
It should be noted that in the leptoquark model these observables deviate significantly from their SM predictions.
\begin{figure}[htb]
\center
\includegraphics[width=7.5cm,height=5.5cm]{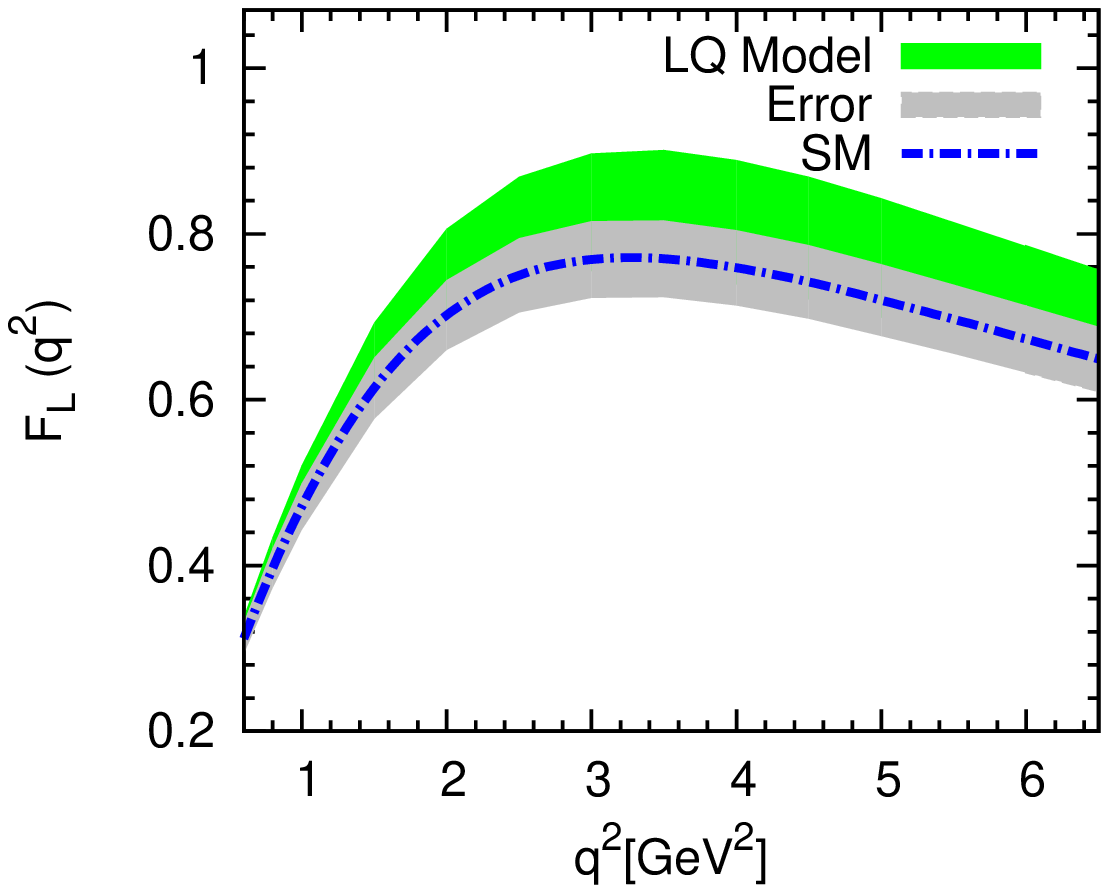}
\quad
\includegraphics[width=7.5cm,height=5.5cm]{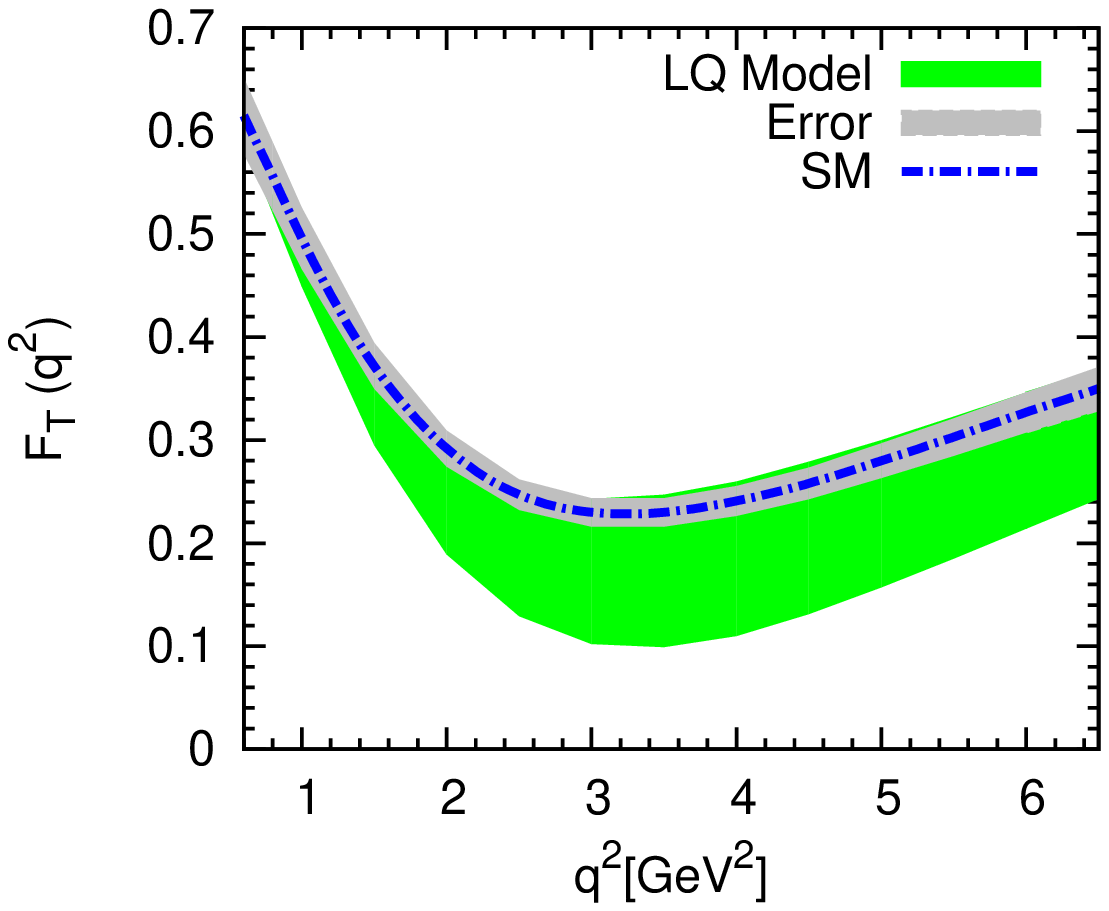}
\includegraphics[width=7.5cm,height=5.5cm]{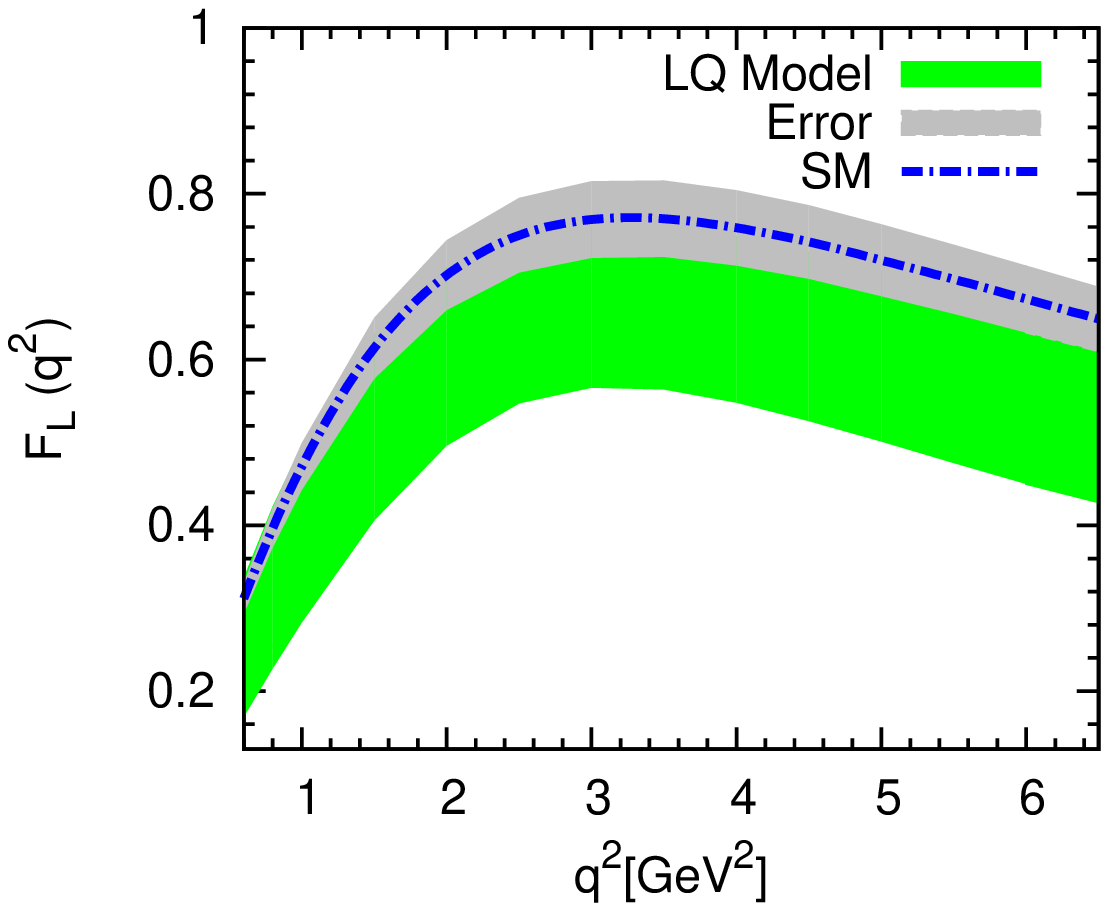}
\quad
\includegraphics[width=7.5cm,height=5.5cm]{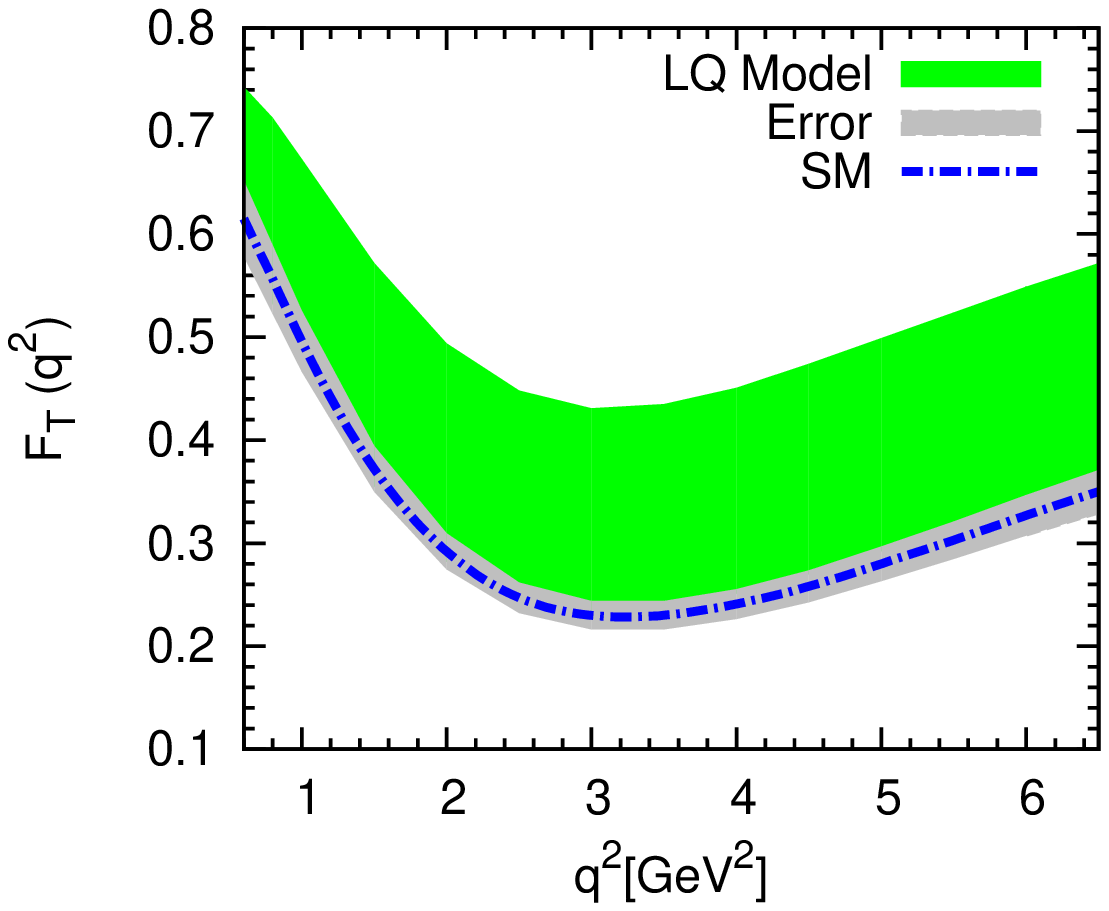}
\caption{Variation of longitudinal  and transverse $K^*$ polarization fractions in the  low $q^2$ region.
The plots in the top panel are for $X(3,2,7/6)$ and those in the bottom panel are for $X(3,2,1/6)$.}
\end{figure}
\begin{figure}[htb]
\centering
\includegraphics[width=7.5 cm,height=5.5 cm]{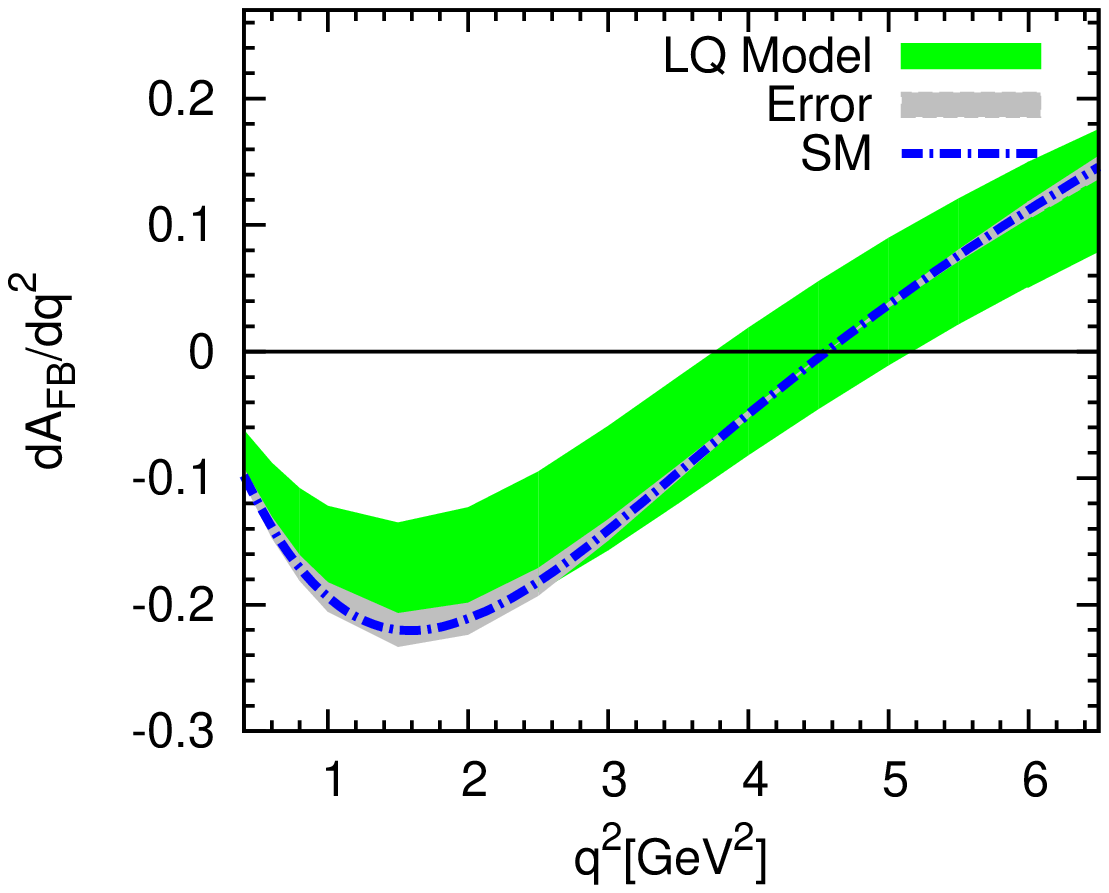}
\quad
\includegraphics[width=7.5cm,height=5.5 cm]{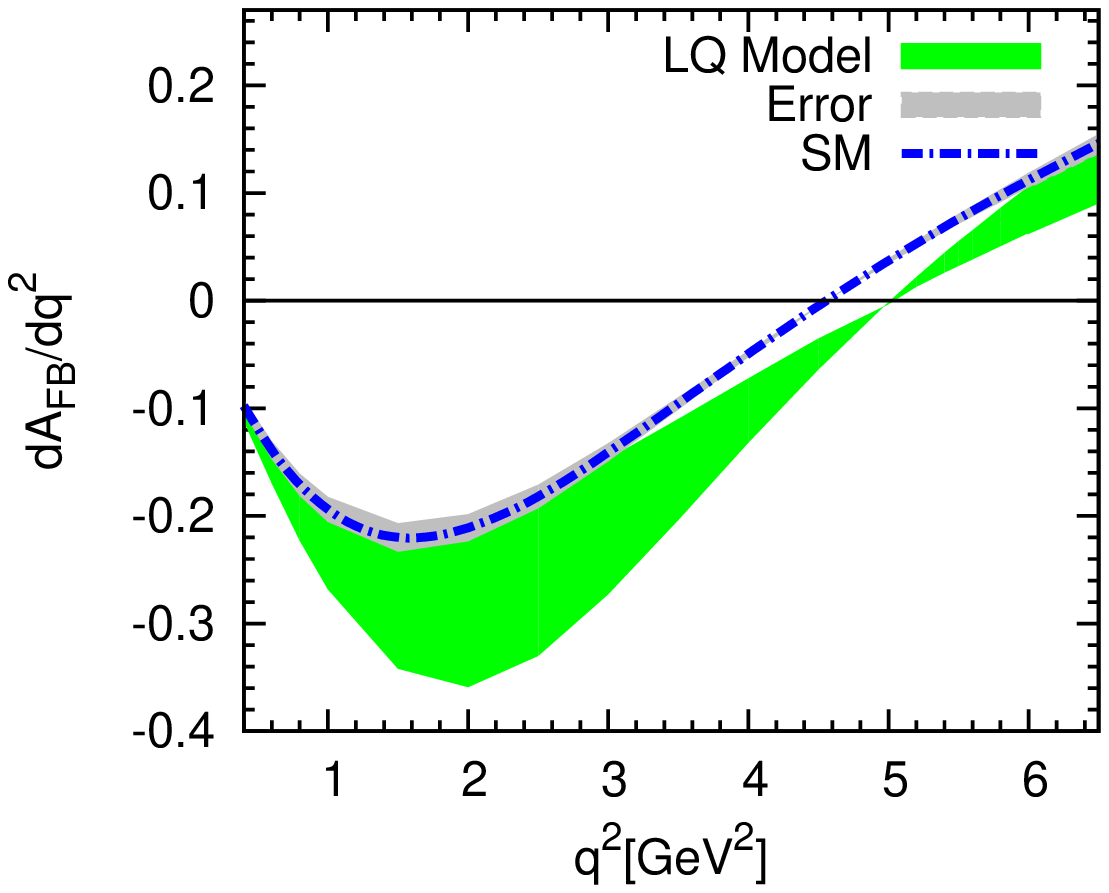}
\caption{The variation of forward backward asymmetry for $X(3,2,7/6)$ (left panel) and for $X(3,2,1/6)$ (right panel) with  $q^2$.}
\end{figure}
\begin{table}[htb]
\caption{The predicted values for the integrated branching ratio, forward-backward asymmetry, isospin asymmetry and the polarisation fractions 
in the range $q^2 \in [1,6]~ {\rm GeV}^2$ for the $B \to K^* \mu^+ \mu^-$ process. }
\begin{center}
\begin{tabular}{c  c  c  c }
\hline
 Observables & SM prediction & Values in LQ model   & Values in LQ model \\
     &  & $X(3,2,7/6)$& $X(3,2,1/6)$ \\
 \hline
 \hline
BR($B \to K^* \mu^+ \mu^-$) & $(7.738 \pm 0.464)\times 10^{-7}$  & $(6.88  \to 8.73 )\times 10^{-7} $ & $(4.8 \to 8.3) \times 10^{-7}$ \\
BR($B \to K^* e^+ e^-$) & $(7.742 \pm 0.465) \times 10^{-7}$ & $(7.03  \to 8.73) \times 10^{-7}$& $(4.85 \to 8.38) \times 10^{-7}$\\
$\langle A_{FB}\rangle$ & $-(0.09 \pm 0.005)$	& $-0.11  \to 0.004$  & $-0.185 \to - 0.08$\\
$\langle A_{I}\rangle$ & $-(0.03 \pm 0.002)$ &  $-0.06 \to -0.04 $ & $-0.02 \to - 0.01$\\
$\langle F_L \rangle$ & 0.71 $\pm$ 0.043 & $0.7 \to 0.8 $ & $ 0.5 \to 0.7$ \\
$\langle F_T\rangle$ & 0.29 $\pm$ 0.017 & $0.2 \to 0.3 $ & $0.3 \to  0.5$  \\
 \hline
\end{tabular}
\end{center}
\end{table}


Another interesting observable is the Isospin asymmetry distribution, which has 
been recently measured by LHCb experiment at 3 fb$^{-1}$ data set  \cite{isospinlhcb}. 
This asymmetry arises due to the non-factorizable part where photon is radiated from  the spectator quark in annihilation and spectator scattering.
The CP-averaged isospin asymmetry is defined as \cite{matias3, lord}
 \begin{equation}
 \frac{dA_I}{dq^2} = \frac{d\Gamma\left[B^0\rightarrow K^{* 0} l^+ l^-\right]/dq^2 
- d\Gamma\left[B^\pm\rightarrow K^{* \pm} l^+ l^-\right]/dq^2}{d\Gamma\left[B^0\rightarrow K^{* 0} l^+ l^-\right]/dq^2 
+ d\Gamma\left[B^\pm\rightarrow K^{* \pm} l^+ l^-\right]/dq^2}\;.
 \end{equation}
 Including longitudinal photon polarizations appearing for $q^2 \neq 0$, the isospin asymmetry distribution in the QCD factorization scheme is given by 
\begin{equation}
A_I\left(q^2\right) = Re\left(b^\perp_d\left(q^2\right) - b^\perp_u\left(q^2\right)\right) \frac{|C_9^{(0) \perp}\left(q^2\right)|^2}{|C_{10}\left(\mu_b\right)|^2 + |C_9^{(0) \perp}\left(q^2\right)|^2} \times \frac{F\left(q^2\right)}{G\left(q^2\right)}
\end{equation}
with 
\begin{equation}
F\left(q^2\right) = 1 + \frac{1}{4} \frac{E^2_{K^*} m^2_B}{q^2 m^2_{K^*}} \frac{\xi^2_\parallel \left(q^2\right)}{\xi^2_\perp \left(q^2\right)} \frac{Re\left(b^\parallel_d \left(q^2\right) - b^\parallel_u \left(q^2\right)\right)}{Re\left(b^\perp_d \left(q^2\right) - b^\perp_u \left(q^2\right)\right)} \frac{|C_9^{(0) \parallel}\left(q^2\right)|^2}{|C_9^{(0) \perp}\left(q^2\right)|^2}
\end{equation}
and 
\begin{equation}
G\left(q^2\right)= 1 + \frac{1}{4} \frac{E^2_{K^*} m^2_B}{q^2 m^2_{K^*}} \frac{\xi^2_\parallel \left(q^2\right)}{\xi^2_\perp \left(q^2\right)} 
\frac{|C_9^{(0) \parallel}\left(q^2\right)|^2 + |C_{10}\left(\mu_b\right)|^2 }{|C_9^{(0) \perp}\left(q^2\right)|^2 + |C_{10}\left(\mu_b\right)|^2 }\;,
\end{equation}
where the generalized standard model Wilson coefficients are  
\begin{equation}
C_9^{(0) \perp}\left(q^2\right) = C_9\left(\mu_b\right) + Y\left(q^2\right) + \frac{2m_b m_B}{q^2} C_7^{eff} \left(\mu_b\right)
\end{equation}
and 
\begin{equation}
C_9^{(0) \parallel}\left(q^2\right) = C_9\left(\mu_b\right) + Y\left(q^2\right) + \frac{2m_b}{m_B} C_7^{eff} \left(\mu_b\right). \hspace{0.8cm}
\end{equation}
The $b^a_q$, $(a=\perp, \parallel)$ terms appearing in the above equations are given as
\begin{equation}
b_q^\perp \left(q^2\right) = \frac{24\pi^2 m_B f_B e_q}{q^2 \xi_\perp \left(q^2\right)C_9^{(0) \perp}\left(q^2\right)} \left( \frac{f^\perp _{K^*}}{m_b} K^\perp_1\left(q^2\right) + \frac{f_{K^*} m_{K^*}}{6\lambda_{B , +}\left(q^2\right)m_B} \frac{K_2^\perp\left(q^2\right)}{1-q^2/m^2_B}\right),
\end{equation}
and
\begin{equation}
b_q^\parallel \left(q^2\right) = \frac{24\pi^2 f_B e_q m_{K^*}}{m_B E_{K^*} \xi_\parallel \left(q^2\right)C_9^{(0) \parallel}\left(q^2\right)} \left( \frac{f _{K^*}}{3\lambda_{B , -}\left(q^2\right)} K^\parallel_1\left(q^2\right)\right). \hspace{4cm}
\end{equation}
 where the expressions for the terms $K_{1,2}^a$ are presented in Appendix E. The variation of isospin asymmetry distribution with respect to dimuon 
invariant mass squared has given in right panel of Fig. 8 and Table III contains the allowed range of isospin asymmetries.
\begin{table}[htb]
\caption{The predicted integrated values of the FFI observables and the CP violating observables in the range $q^2 \in [1,6]~{\rm  GeV}^2$ for the $B \to K^* \mu^+ \mu^-$ process. }
\begin{center}
\begin{tabular}{ c c c c }
\hline
Observables & SM prediction & Values in LQ model   & Values in LQ model \\
     &  & $X(3,2,7/6)$ & $X(3,2,1/6)$ \\
 \hline
 \hline
$\langle P_1 \rangle$ & $-0.044 \pm 0.003$ & $-0.037 \to -0.046$  & $-0.017 \to 0.14$\\
$\langle P_2 \rangle$ & 0.203 $\pm$ 0.012 & 0.08 $\to$ 0.23  & 0.19 $\to$ 0.21\\
$\langle P_3 \rangle$ & $-(6.0 \pm 0.4)\times 10^{-4}$	~& ~$-(9.4 \to 1.8)\times 10^{-3}$   & $-0.014 \to 0.08 $\\  
$\langle P_4 \rangle$ & 0.395 $\pm$ 0.024 & 0.294 $\to$ 0.45  & 0.24 $\to$ 0.52 \\ 
$\langle P_5 \rangle$ & $-0.204 \pm 0.012$ & $-0.42 \to - 0.13$  & $-0.39 \to -0.13$\\
$\langle P_6 \rangle$ & $-0.0075 \pm 0.0005 $ & $ -0.07 \to 0.08 $ & 0.079 $\to$ 0.112 \\
 $\langle A_T^{(3)} \rangle$ & 0.55 $\pm$ 0.033 & 0.56 $\to$ 0.6 & 0.37 $\to$ 0.6 \\
 $\langle A_T^{(4)} \rangle$ & 0.87 $\pm$ 0.05 & 0.82 $\to$ 0.94 & 0.99 $\to$ 1.5\\
 $\langle A_{Im} \rangle$ & $( 1.7 \pm 0.1)\times 10^{-4}$ & $(2.0 \to 5.3)\times 10^{-4}$ & -0.023 $\to$ 0.005\\
 $\langle \alpha_{K^*} \rangle$ & 3.73 $\pm$ 0.22 & 3.6 $\to$ 5.6 &  1.8 $\to$ 3.6\\
 
 \hline
\end{tabular}
\end{center}
\end{table}

\begin{figure}[htb]
\centering
\includegraphics[width=7.5 cm,height=5.5 cm]{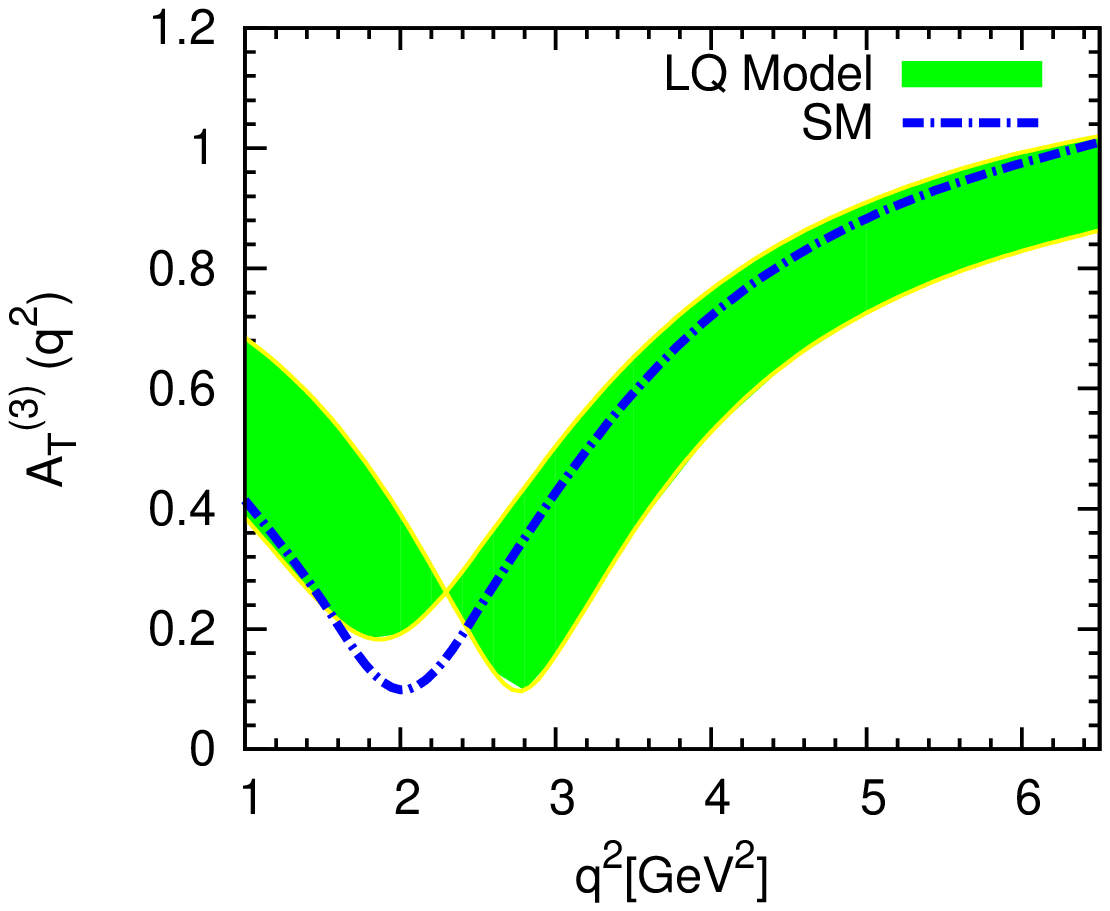}
\quad
\includegraphics[width=7.5 cm,height=5.5 cm]{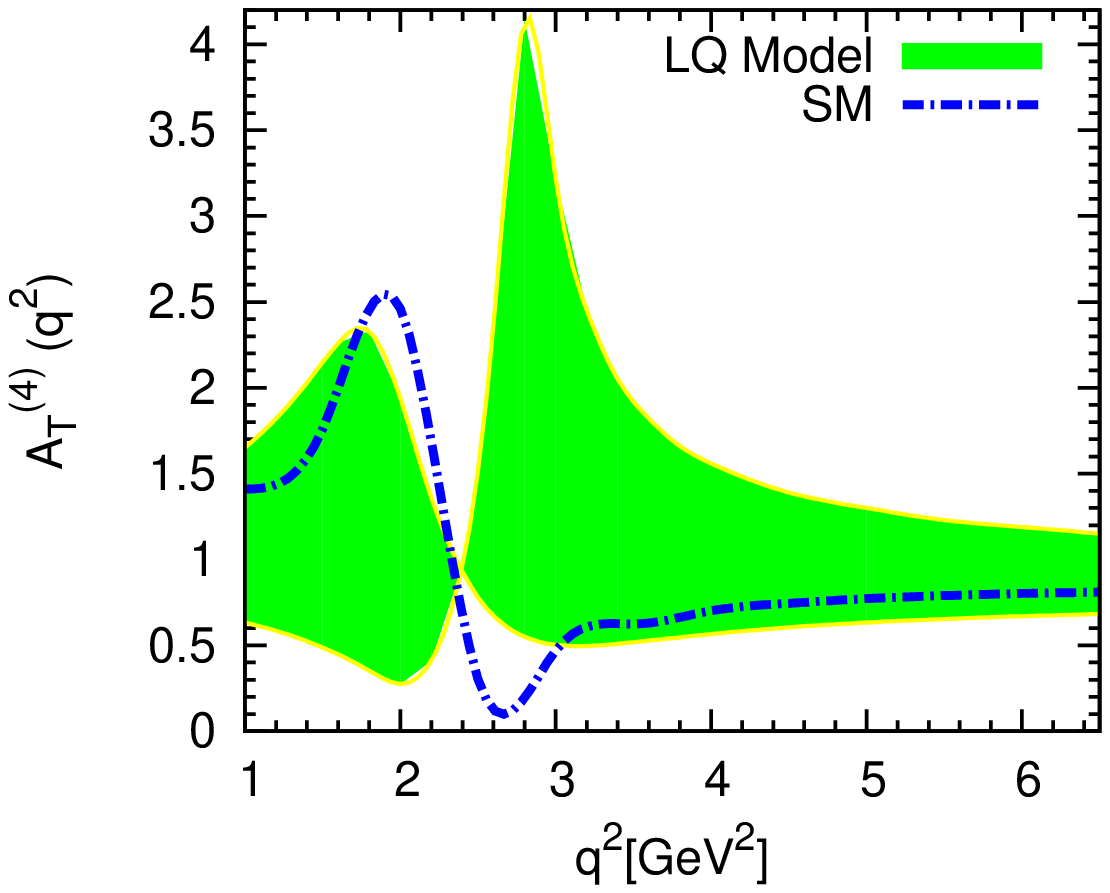}
\includegraphics[width=7.5cm,height=5.5cm]{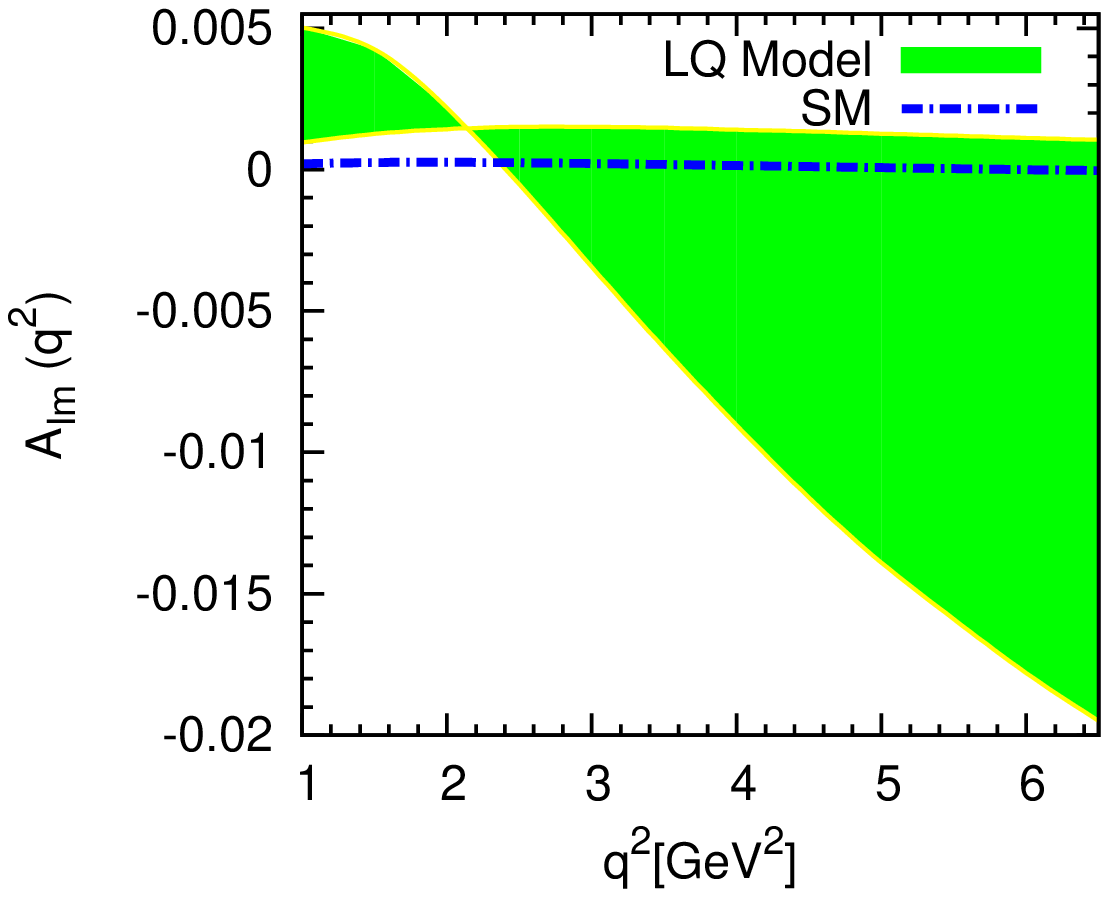}
\quad
\includegraphics[width=7.5cm,height=5.5cm]{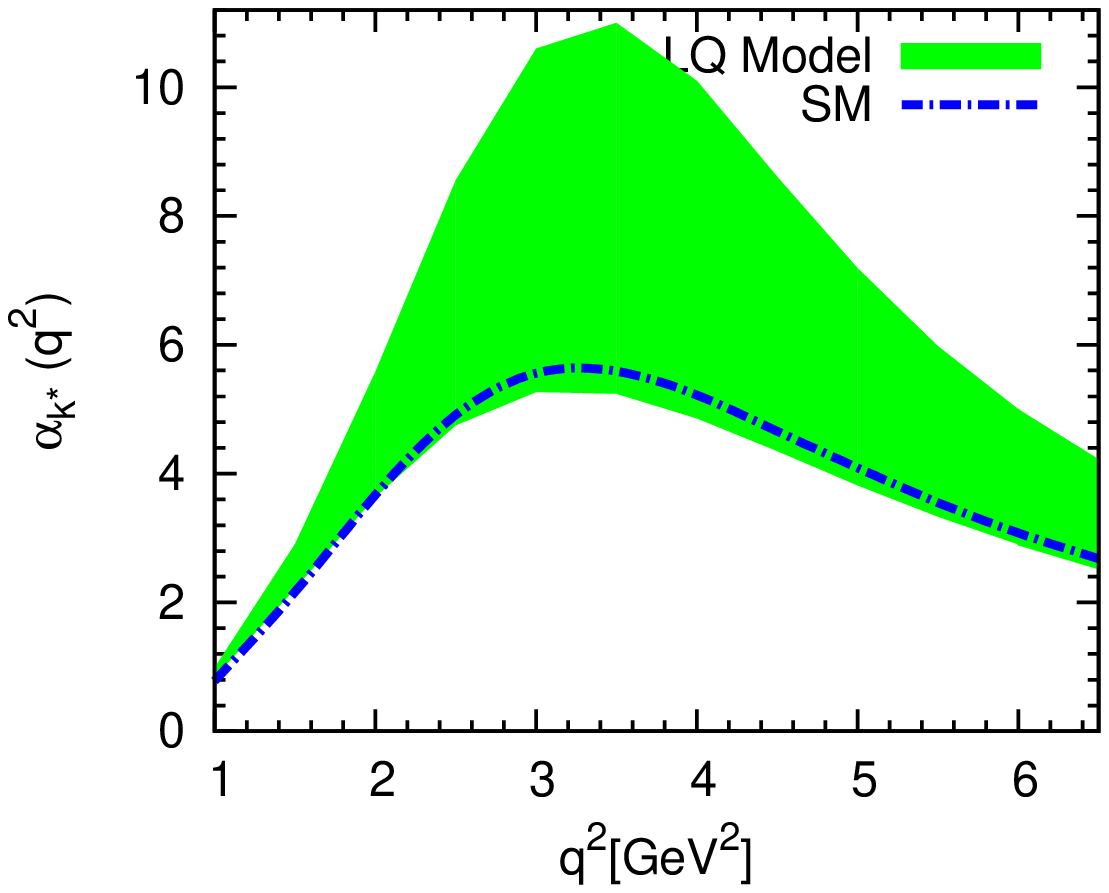}
\caption{The variation of transverse asymmetry parameters $A_T^{(3)}$, $A_T^{(4)}$, $A_{\rm Im}$  and the $K^*$ polarization factor  
 with $q^2$ in high recoil for $X(3,2,7/6)$ LQ.}
\end{figure}
\begin{figure}[htb]
\centering
\includegraphics[width=7.5cm,height=5.5cm]{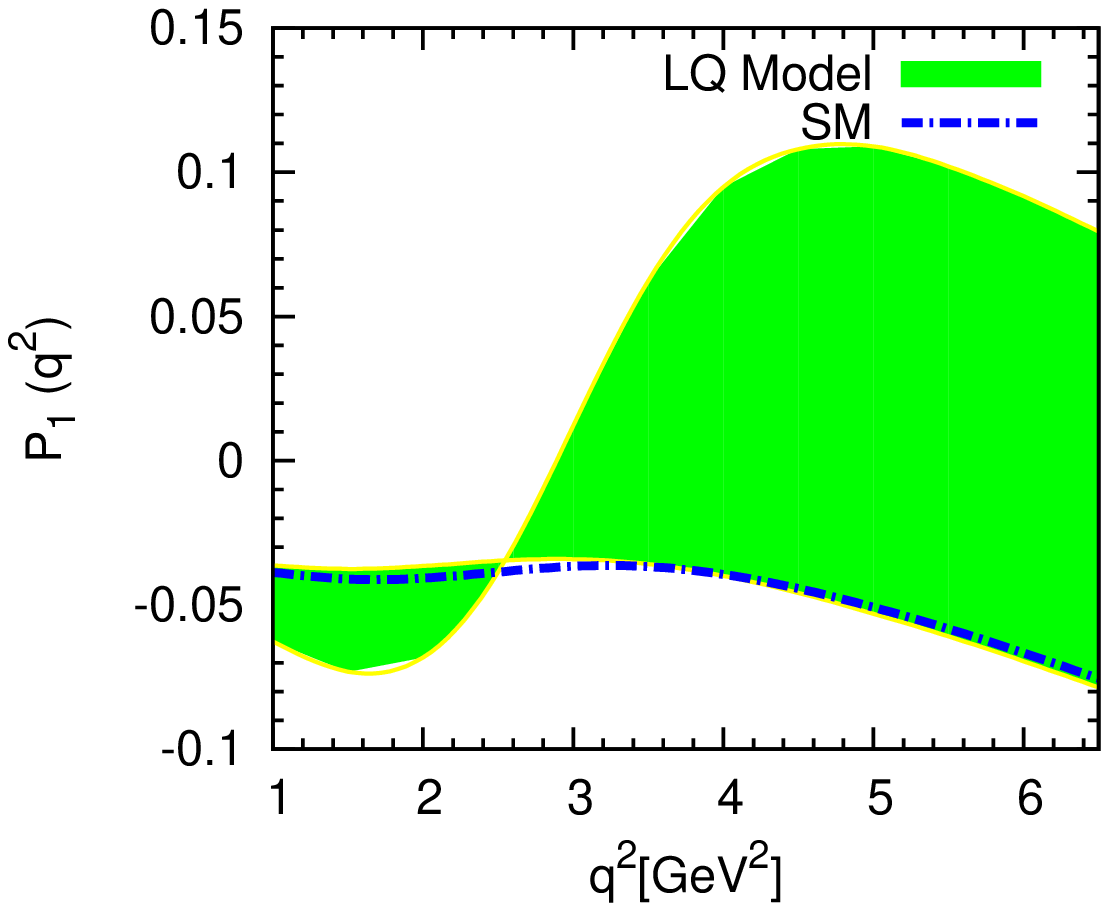}
\quad
\includegraphics[width=7.5cm,height=5.5cm]{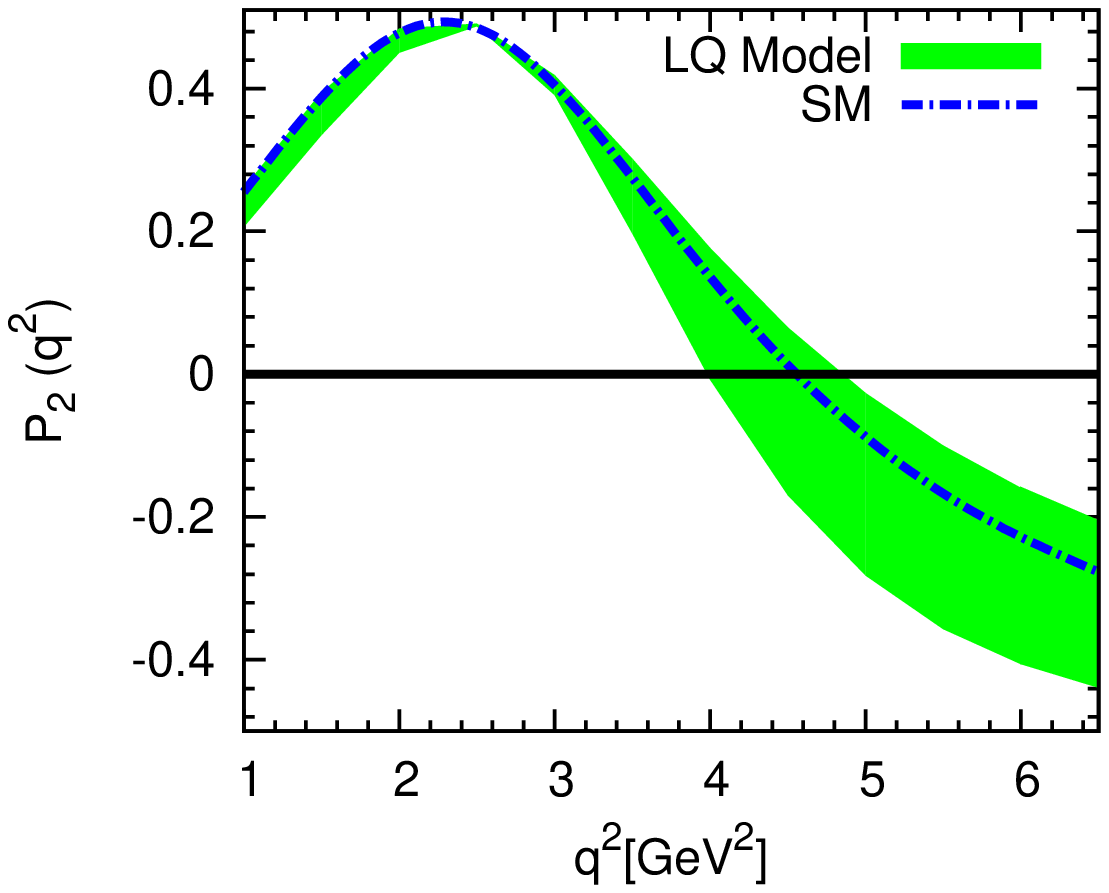}
\quad
\includegraphics[width=7.5cm,height=5.5cm]{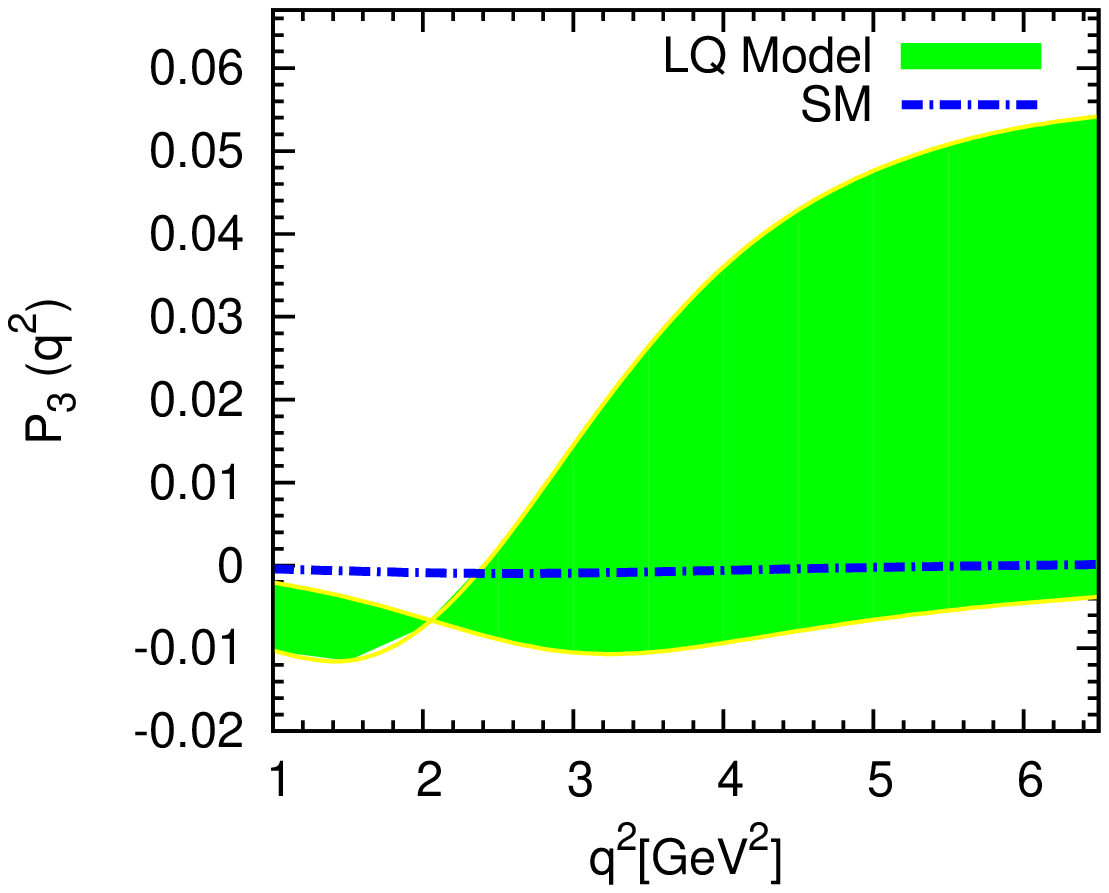}
\quad
\includegraphics[width=7.5cm,height=5.5cm]{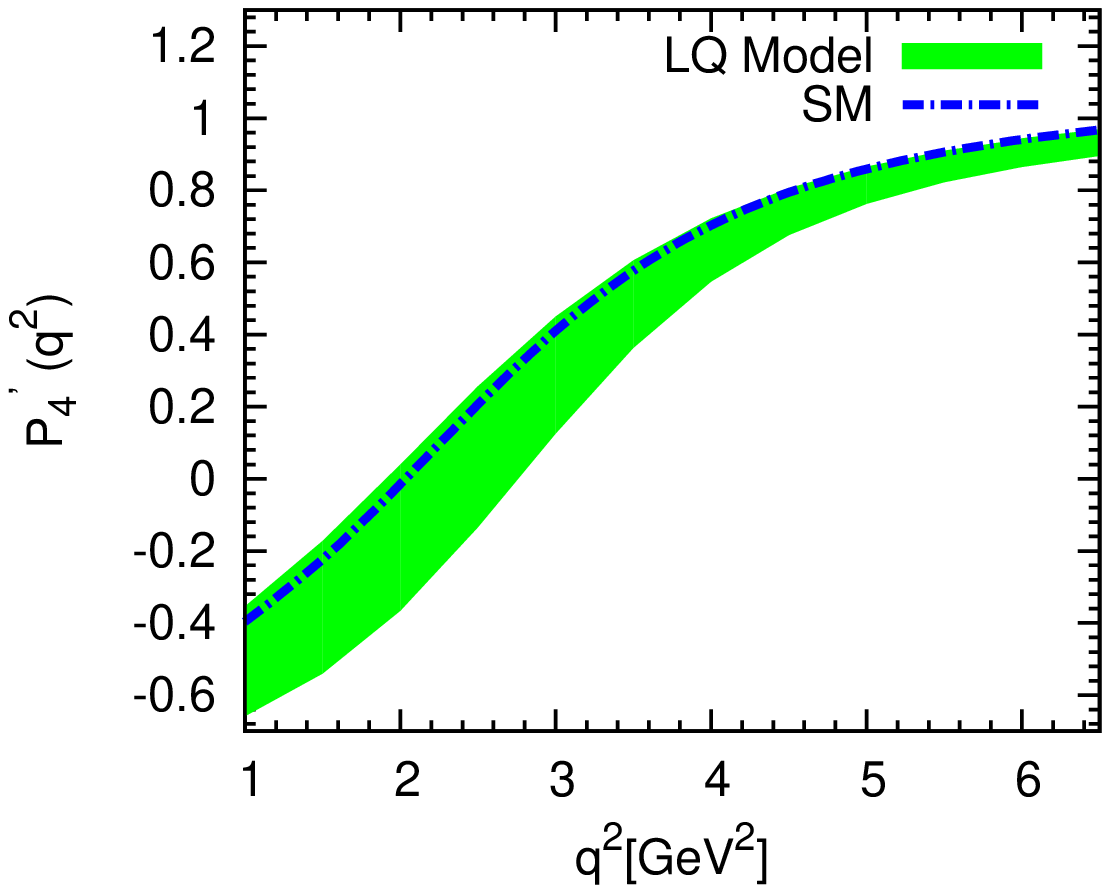}
\quad
\includegraphics[width=7.5cm,height=5.5cm]{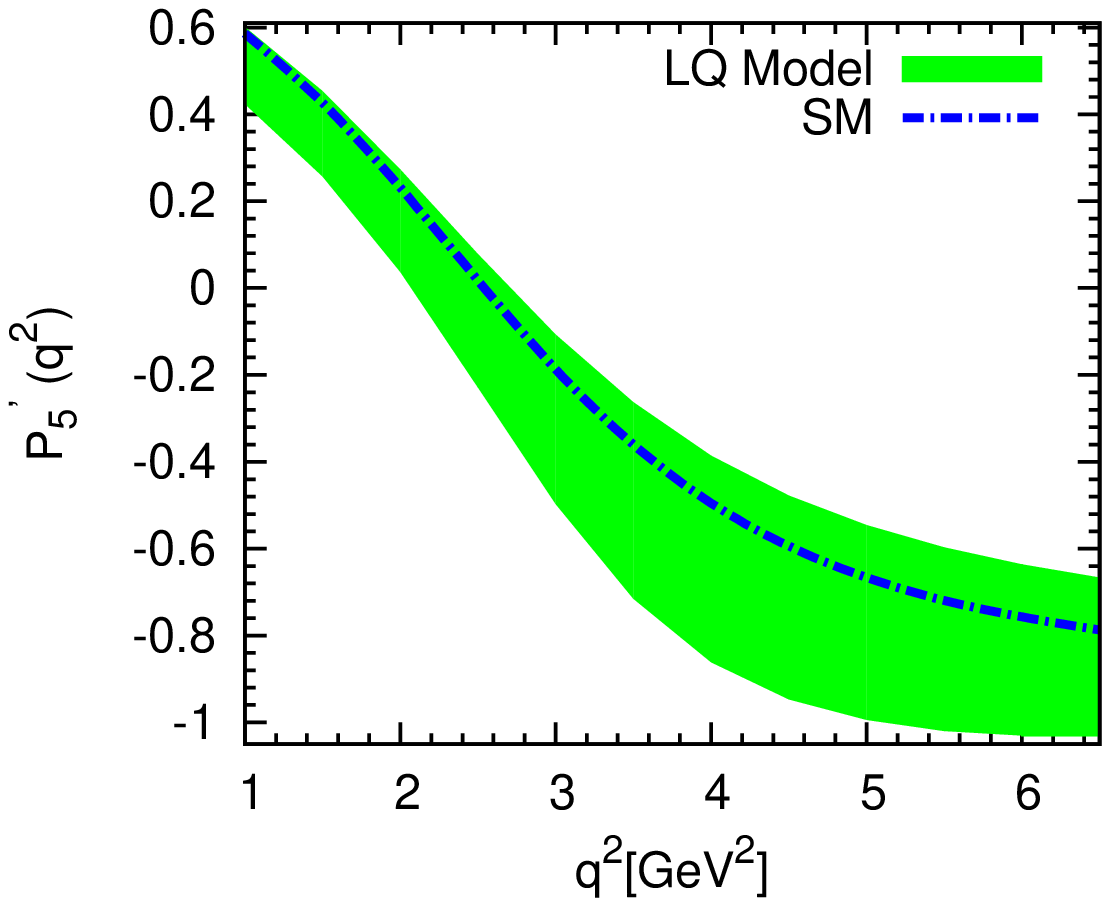}
\quad
\includegraphics[width=7.5cm,height=5.5cm]{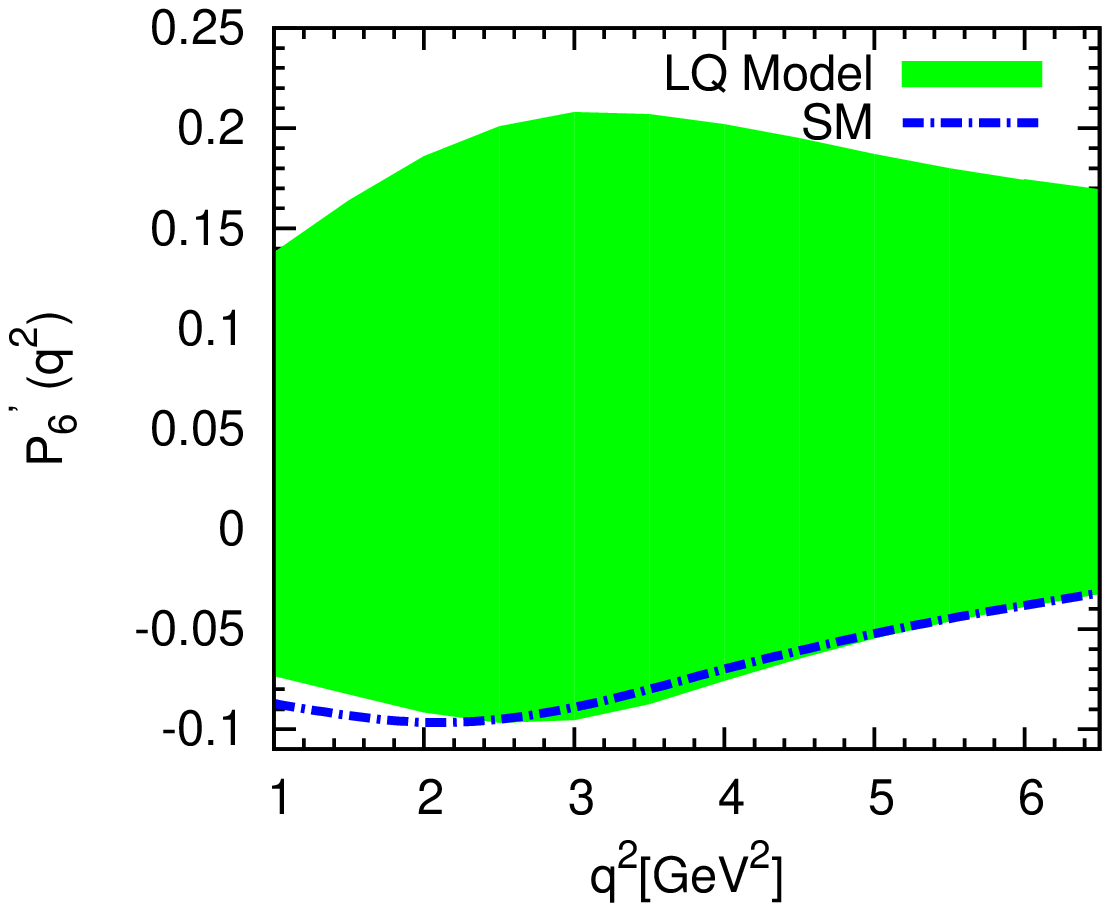}
\caption{The variation of the  observables $P_{1,2,3}$ and $P_{4,5,6}'$ with  $q^2$ for $X(3,2,7/6)$ LQ.}
\end{figure}
\begin{table}[htb]
\caption{The predicted values for the integrated branching ratio, forward-backward asymmetry and isospin asymmetry  in the range $q^2 \in [14.2, 19]~ {\rm GeV}^2$ 
for the $B \to K^* \mu^+ \mu^-$ process. }
\begin{center}
\begin{tabular}{c c c }
\hline
\hline
 Observables & SM prediction & Values in LQ model $X(3,2,7/6)$ \\
 \hline
BR($B \to K^* \mu^+ \mu^-$) & $(8.5 \pm 0.51)\times 10^{-7}$ & $(8.93 \to  9.31 )\times 10^{-7}$\\
BR($B \to K^* e^+ e^-$) & $(8.52 \pm 0.511) \times 10^{-7}$ & $(8.92 \to 9.32)\times 10^{-7}$ \\
$\langle A_{FB}\rangle$ & 0.4	& 0.34 $\to$ 0.38 \\
$\langle A_{I}\rangle$ & $-(2.75 \pm 0.17)\times10^{-3}$ & $(-3.3 \to 2.5  )\times 10^{-3}$ \\
$\langle P_2 \rangle$ & $-0.42$ & $-0.41 \to -0.38 $\\
$\langle A_T^{(4)}\rangle$ & 0.57 & 0.53 $\to$ 0.58 \\
$\langle F_L \rangle$ & $0.35$ & $0.35 $\\
$\langle F_T\rangle$ & 0.65 & 0.65 $\to$ 0.66 \\
 \hline
 \hline
\end{tabular}
\end{center}
\end{table}

\begin{figure}[htb]
\centering
\includegraphics[width=7.5 cm,height=5.5 cm]{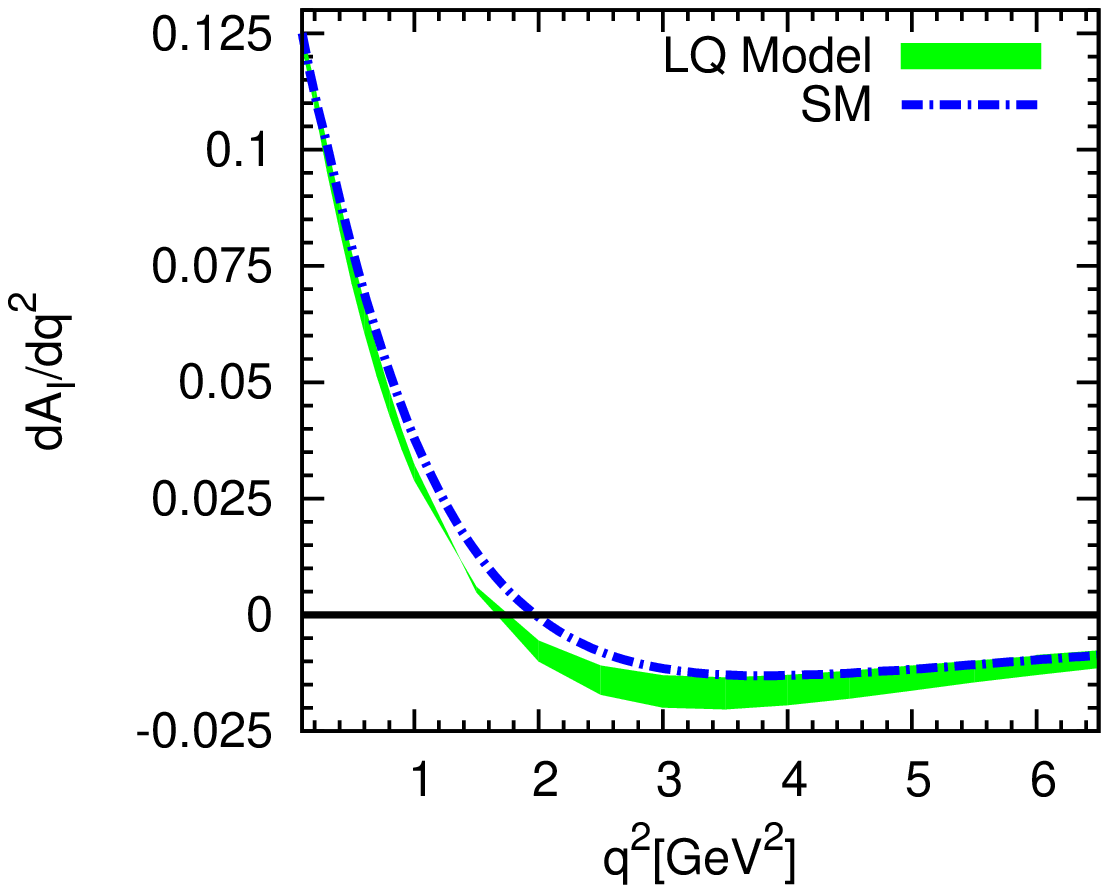}
\quad
\includegraphics[width=7.5cm,height=5.5 cm]{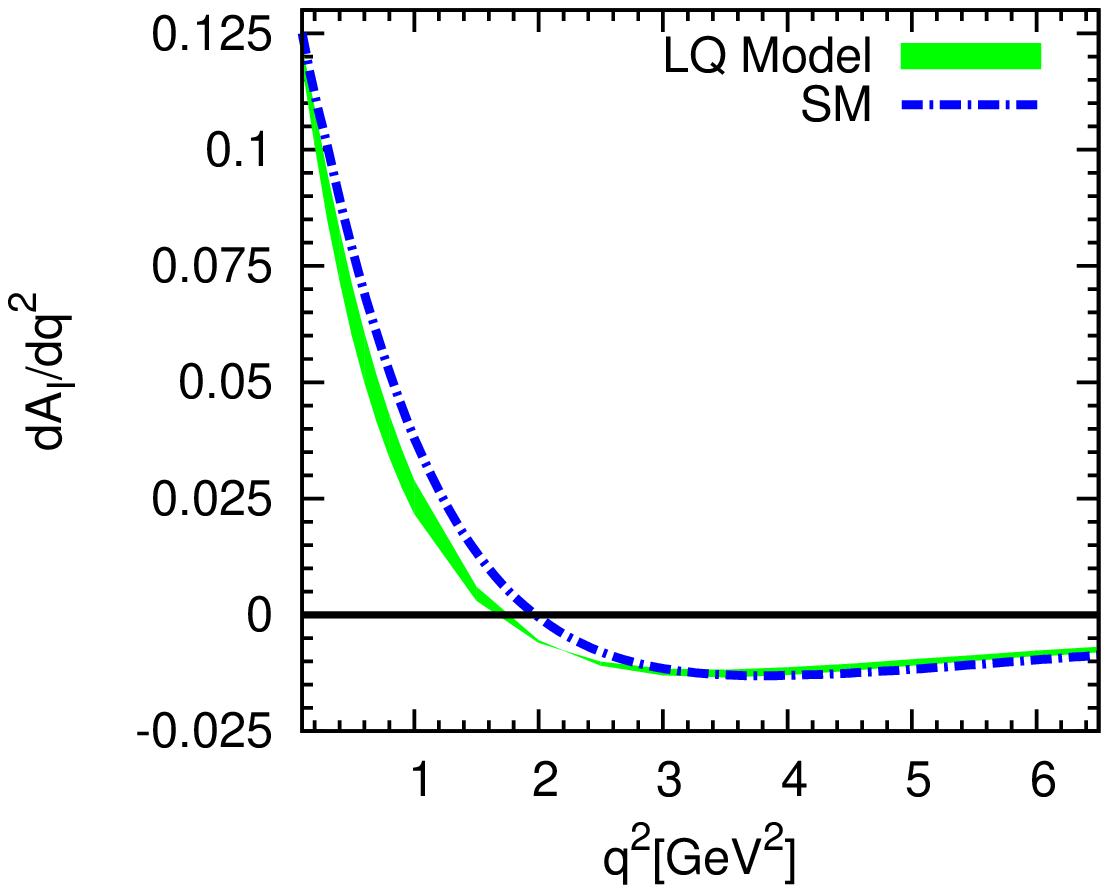}
\caption{The variation of isospin asymmetry for $X(3,2,7/6)$ (left panel) and for $X(3,2,1/6)$ (right panel) with  $q^2$.}
\end{figure}
\subsection{Observables in the low recoil}
At low recoil the exclusive $\bar{B} \rightarrow \bar{K}^* l^+ l^-$ decays depend on the improved form factor relations and an operator product 
expansion (OPE) in $1/Q$. The OPE controls the non-perturbative contributions from four-quark operators and is important for charm quark, whose operators 
 enter without any suppression from CKM matrix elements and small Wilson coefficients.
The QCD operator identity for massless strange quark $\left(m_s = 0\right)$ is \cite{dyk2, dan}
\begin{equation}
i\partial^\nu \left(\bar{s}i\sigma_{\mu \nu}b\right) = -m_b \left(\bar{s}\gamma_\mu b \right) 
+ i\partial_\mu \left(\bar{s}b\right) - 2\left(\bar{s}i\overleftarrow{D}_\mu b\right)\;,
\end{equation}
which allows to extract relation between the form factors $T_1$ and $V$ and the matrix elements of the current $\bar{s}i\overleftarrow{D}_\mu b$.
The improved Isgur-Wise relations to leading order in $1/m_b$ including radiative corrections are 
\beqa
T_1\left(q^2\right) &=& \kappa V\left(q^2\right)\;,\nn\\
T_2\left(q^2\right) &=& \kappa A_1\left(q^2\right)\;,\nn\\
 T_3\left(q^2\right) & = & \kappa A_2\left(q^2\right)\frac{m^2_B}{q^2}\;,
\eeqa
where 
\begin{equation}
\kappa = \left( 1+\frac{2D_0^{(\nu)}\left(\mu\right)}{C_0^{\left(\nu\right)}\left(\mu\right)}\right) \frac{m_b\left(\mu\right)}{m_B}\;.
\end{equation}
and at $\mu = m_b$ including $\mathcal{O}(\alpha_s)$ corrections, it reads $\kappa = 1+ \mathcal{O}(\alpha^2_s)$.
We have shown the variation of branching ratio for $\bar{B} \rightarrow \bar{K}^* \mu^+ \mu^-$ (left panel) and $\bar{B} 
\rightarrow \bar{K}^* e^+ e^-$ (right panel) with $q^2$ in Fig. 9. In the low recoil region the variation of forward-backward asymmetry, 
 isospin asymmetry, longitudinal and transverse polarization fractions of $K^*$  with respect to $q^2$ are shown in Fig. 10. Fig. 11 
shows the variation of  $P_2$, $A_T^4$  with respect to dimuon invariant mass squared. 
Table V contains the integrated values of branching ratio, forward-backward asymmetry and isospin asymmetry in the low recoil 
\textit{i.e.} $q^2 \in [14.2, 19]~{\rm GeV}^2$. It should be noted from these figures that at high $q^2$, there is no significant deviation between
the SM  results and the leptoquark predictions.  
\begin{figure}[htb]
\centering
\includegraphics[width=7cm,height=5cm]{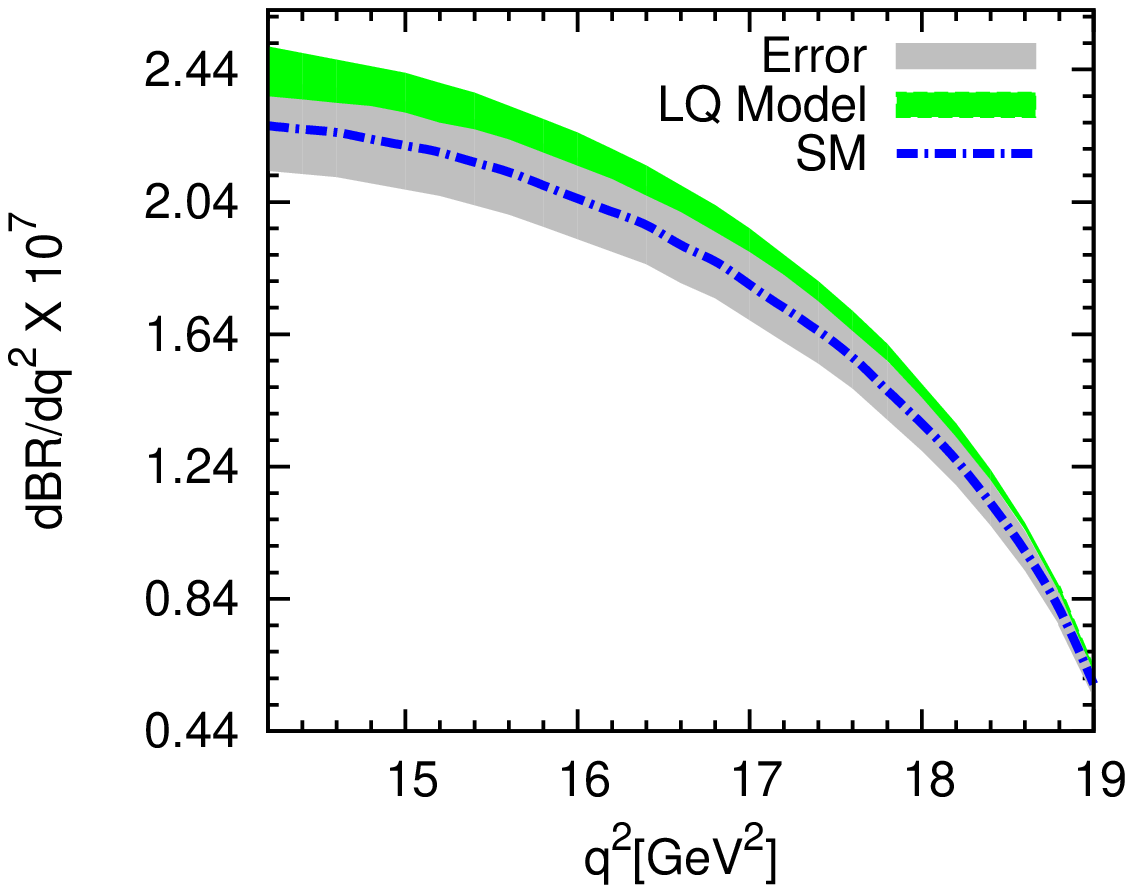}
\quad
\includegraphics[width=7cm,height=5cm]{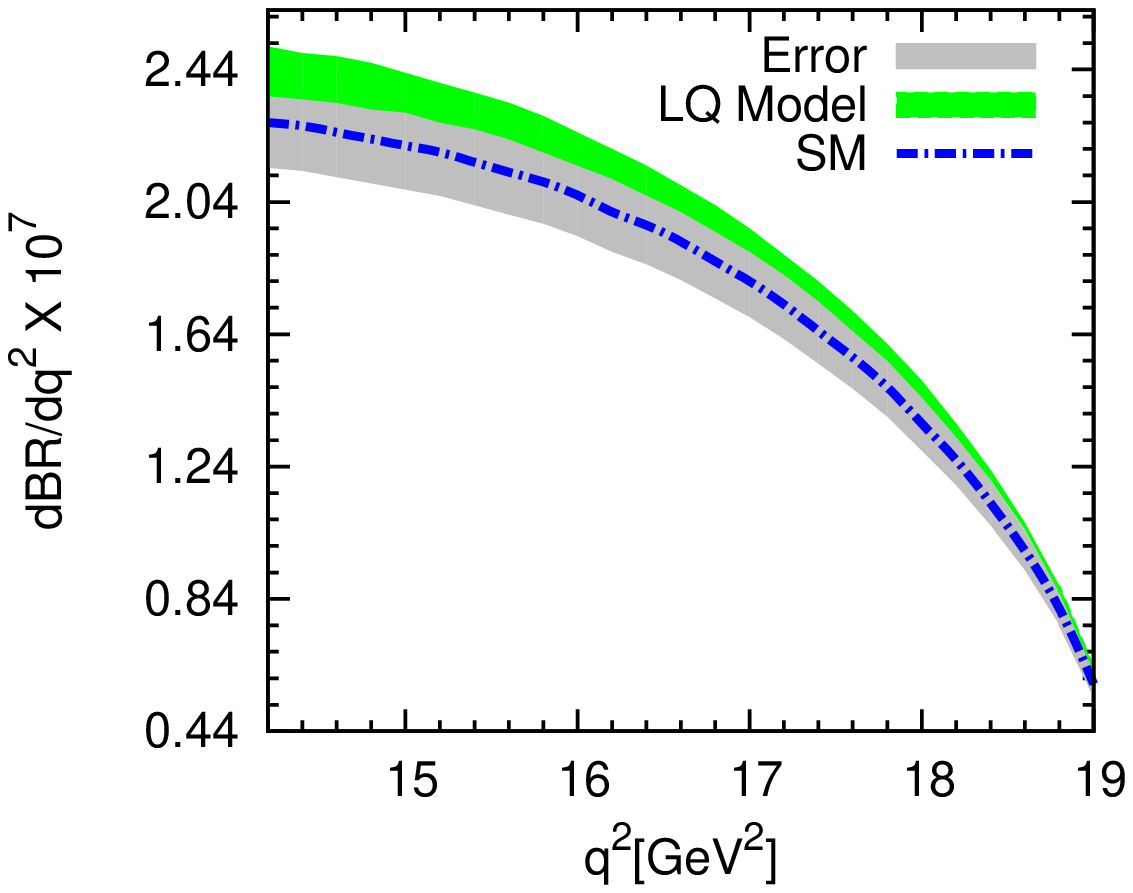}
\caption{The variation of branching ratio for $\bar{B} \rightarrow \bar{K}^* l^+ l^-$ with high $q^2$ for $X(3,2,7/6)$ LQ.}
\end{figure}
\begin{figure}[htb]
\centering
\includegraphics[width=7cm,height=5cm]{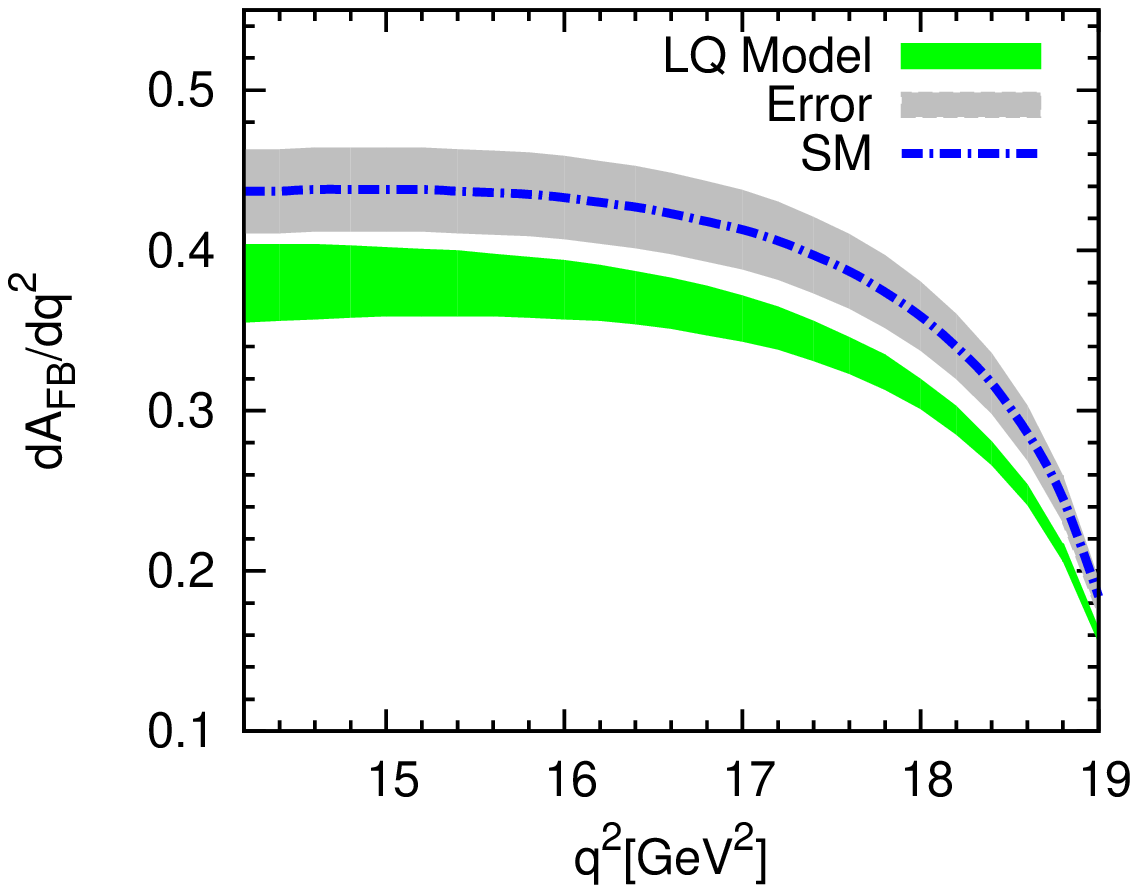}
\quad
\includegraphics[width=7.5cm,height=5cm]{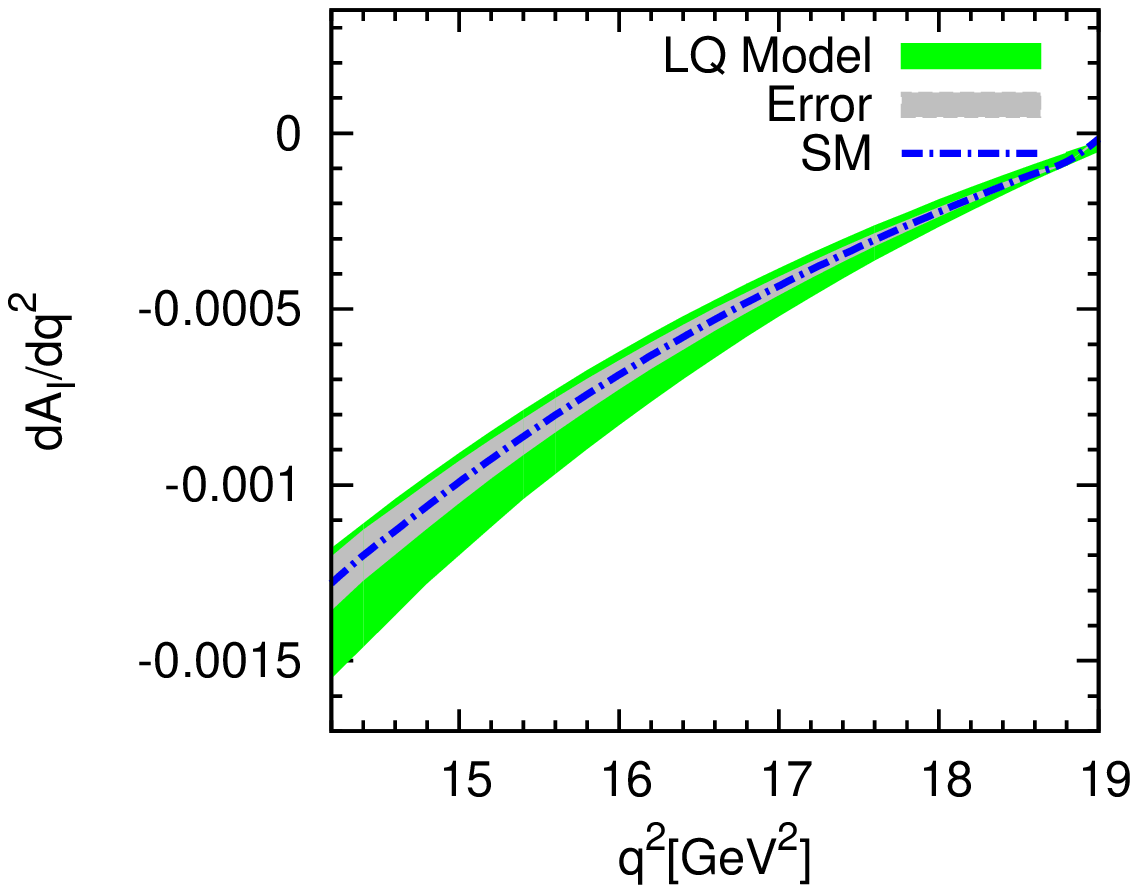}
\includegraphics[width=7cm,height=5cm]{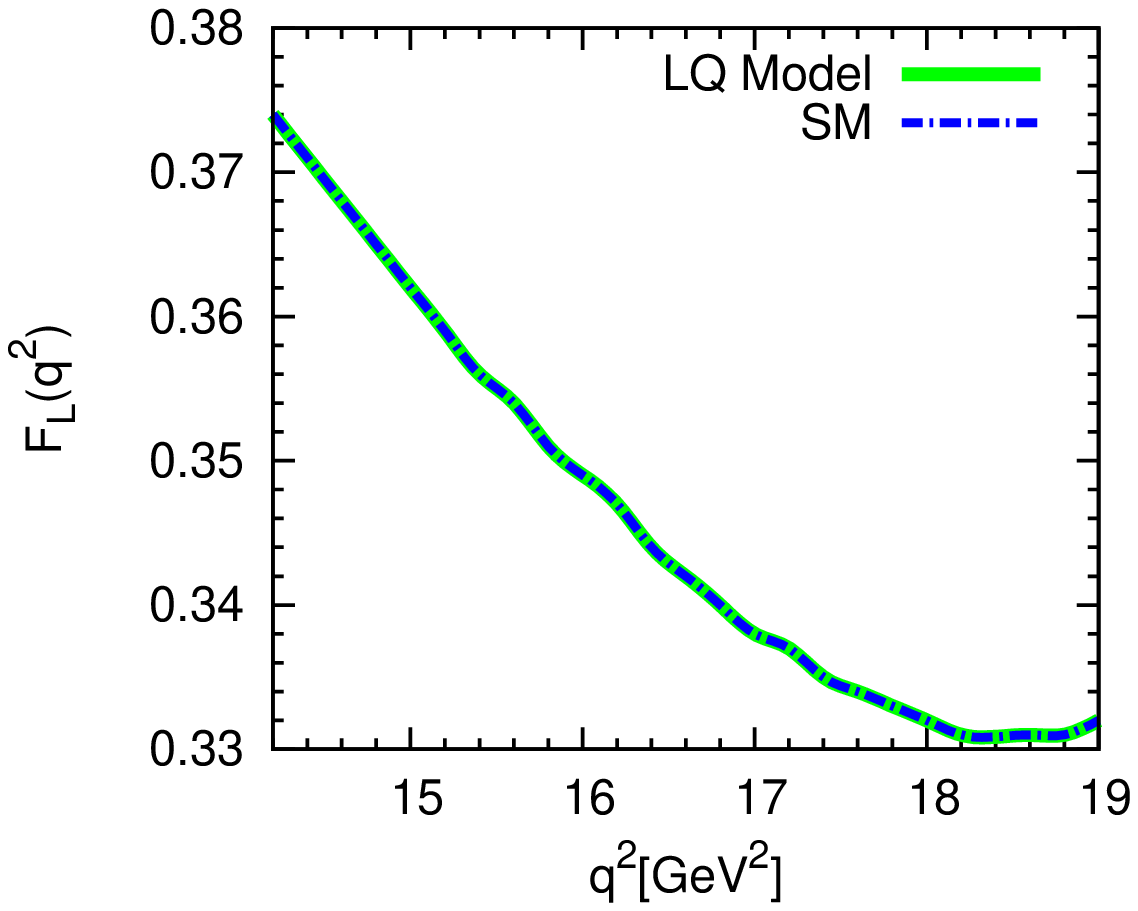}
\quad
\includegraphics[width=7cm,height=5cm]{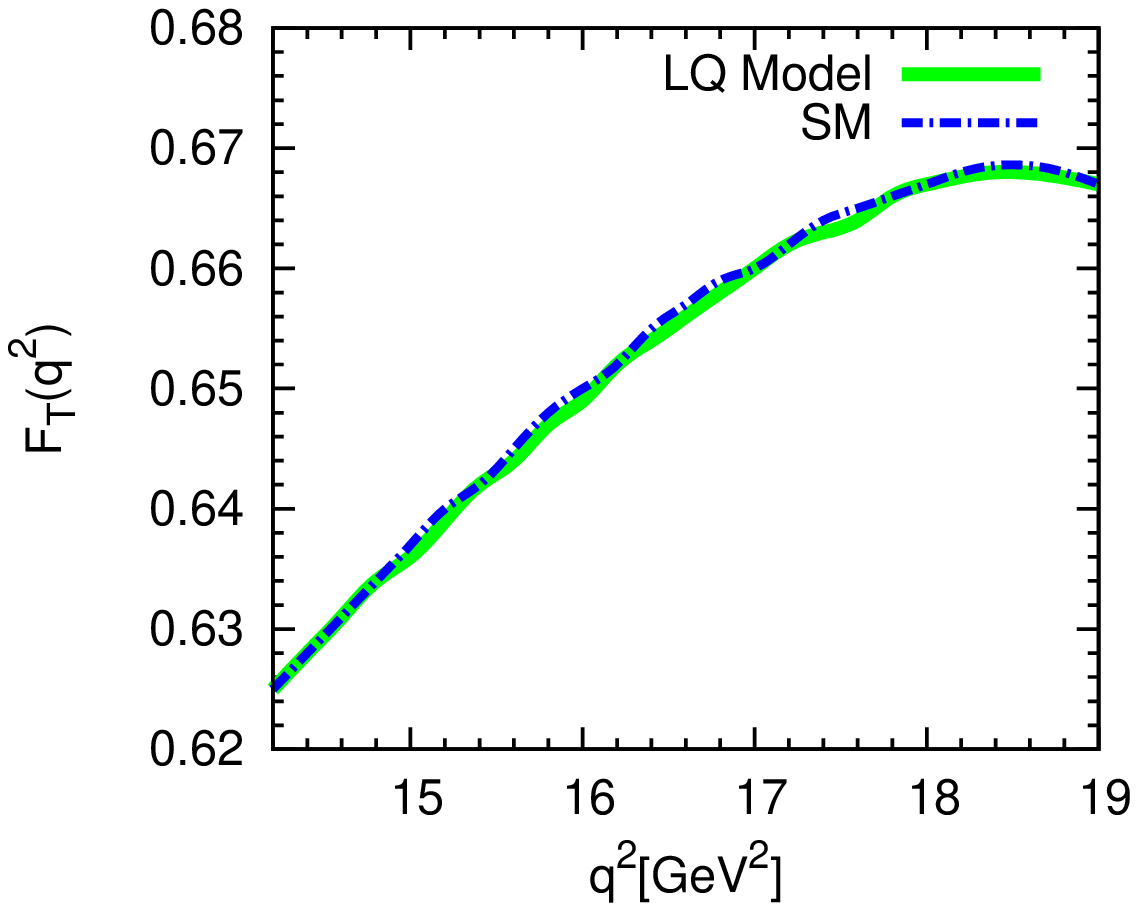}
\caption{The variation of forward backward asymmetry,  isospin asymmetry,  longitudinal and transverse  polarization fraction with high $q^2$
for $X(3,2,7/6)$ LQ.}
\end{figure}

\begin{figure}[htb]
\centering
\includegraphics[width=7cm,height=5cm]{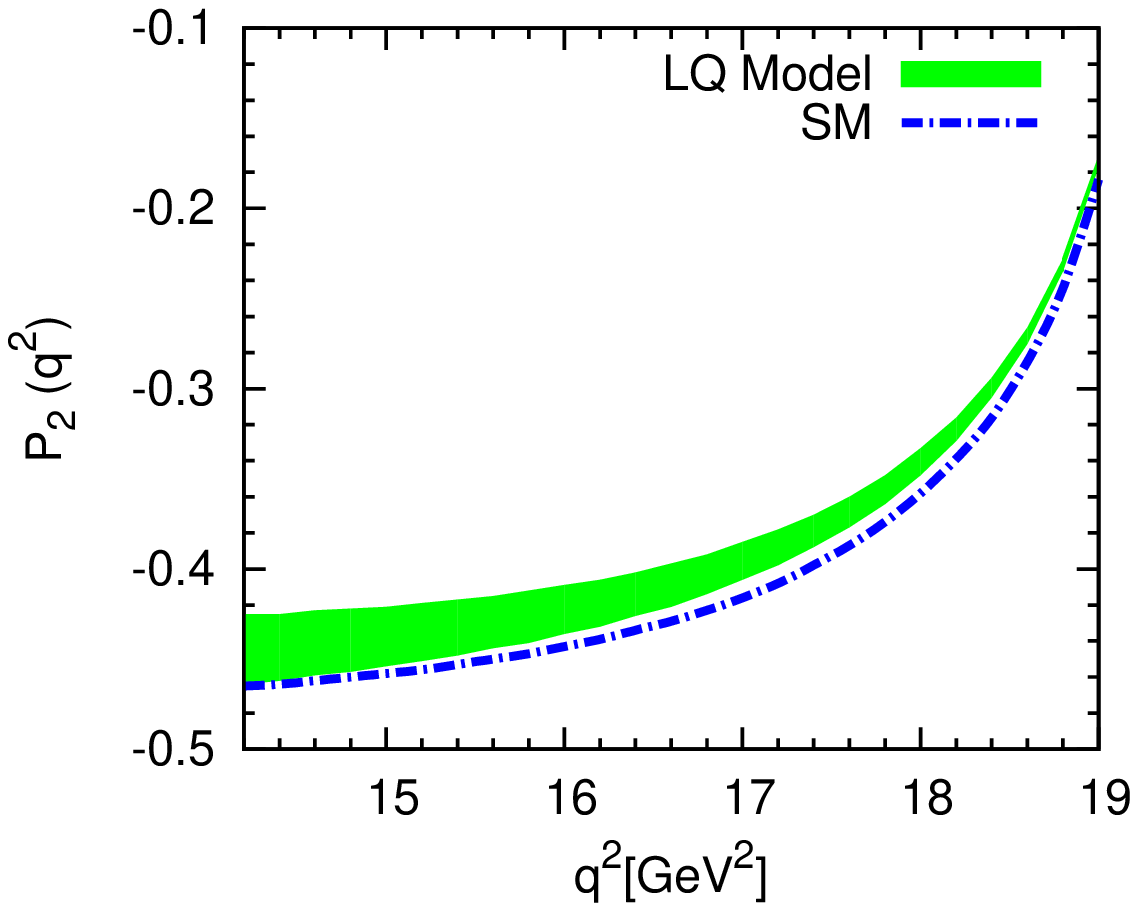}
\quad
\includegraphics[width=7cm,height=5cm]{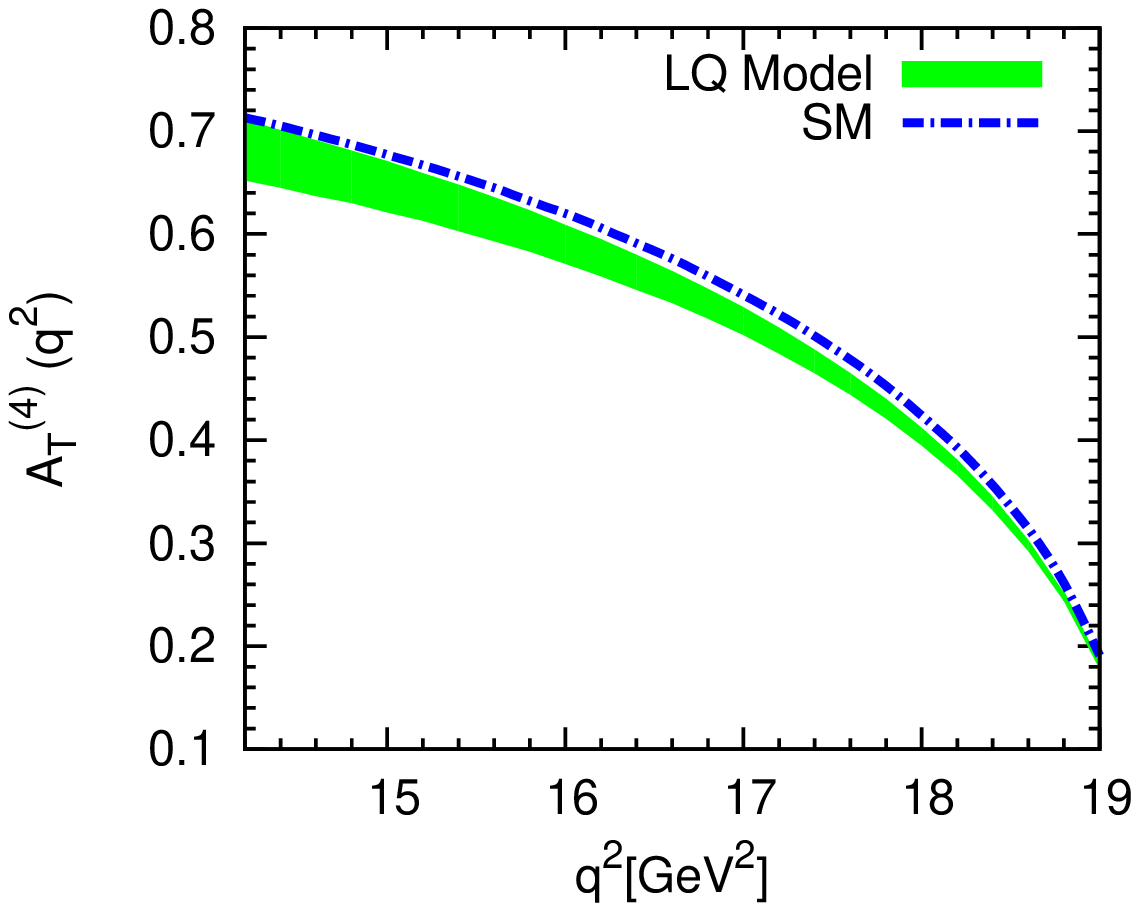}
\caption{Variation of the observables $P_{2}$, and $A_T^{(4)}$ for $X(3,2,7/6)$ LQ.}
\end{figure}

\section{Conclusion}
In this paper we have studied the rare semileptonic decays $\bar{B}_d^0 \to \bar{K}^* l^+ l^-$  using the simple re-normalizable leptoquark 
model in which a single scalar leptoquark  is added to the
standard model with the requirement that proton decay would not be
induced in perturbation theory. The leptoquark parameter space is constrained  using the recent measurement on $B_s \rightarrow \mu^+ \mu^- $.
Using such parameter space we obtained the bounds on the product of leptoquark couplings. We then estimated the branching ratios, 
isospin asymmetries and forward-backward asymmetries for $\bar{B} \rightarrow \bar{K}^* l^+ l^-$ process in full physical region except 
the intermediate region of $q^2$. 
The CP violating observables and the form factor independent observables have also been studied in the leptoquark model.
It is found that  in these models, there could be  significant deviations in these observables  in high recoil but comparatively less in low recoil
regimes. We  found that the time-integrated values of some of the asymmetry parameter  have deviated significantly from 
their corresponding SM values, the observation of which in the LHCb experiment would provide the possible existence of leptoquarks.
\\

{\bf Acknowledgments} 

We would like to thank Science and Engineering Research Board (SERB),
Government of India for financial support through grant No. SB/S2/HEP-017/2013.

\appendix
\section{Transversity amplitudes at NLO in the large recoil}
In the large recoil limit the transversity amplitudes at next to leading order (NLO) within QCDf can be given as \cite{hiller2, ball}
\begin{eqnarray}
A^{L,R}_\perp &=& N\sqrt{2 \lambda} \left[ \left( (C_9+ C_9^{NP}+C_9'^{NP})\mp (C_{10}+C_{10}^{NP}+C_{10}'^{NP}\right) 
\frac{V\left(q^2\right)}{m_B + m_{K^*}} + \frac{2m_b}{q^2}\mathcal{T}_\perp^+ \right],\nn\\
 A_\parallel^{L,R} & = &  -N\sqrt{2}\left(m^2_B - m^2_{K^*}\right) \Bigg[ \left((C_9+C_9^{NP}-C_9'^{NP})\mp (C_{10}+C_{10}^{NP}-C_{10}'^{NP})
\right) \frac{A_1\left(q^2\right)}
{m_B - m_{K^*}}\nn\\
 &&\hspace*{2.5 true cm}+ \frac{4m_b}{m_B} \frac{E_{K^*}}{q^2}\mathcal{T}_\perp^- \Bigg], \nn\\
  A_0^{L,R} & = &  - \frac{N}{2m_{K^*}\sqrt{q^2}} \Bigg[ \left((C_9 +C_9^{NP}-C_9'^{NP}) \mp (C_{10}+ C_{10}^{NP}-C_{10}'^{NP}) \right)\nn\\
&\times &\left[ \left( m^2_B - m^2_{K^*} - q^2 \right) 
\left(m_B + m_{K^*}\right)A_1\left(q^2\right) - \lambda \frac{A_2\left(q^2\right)}{m_B + m_{K^*}}\right] \nn\\
& + &2m_b \left[\frac{2E_{K^*}}{m_B}
\left( m^2_B + 3m^2_{K^*}-q^2\right) \mathcal{T}_\perp^- - \frac{\lambda}{m^2_B - m^2_{K^*}}\left(\mathcal{T}_\perp^- 
+ \mathcal{T}_\parallel^-\right) \right]\Bigg]\;,\nn\\
 A_t &=& \frac{2N}{\sqrt{q^2}} \frac{E_{K^*}}{m_{K^*}}\frac{\xi_\parallel}{\Delta_\parallel}\sqrt{\lambda}
\left (C_{10}+C_{10}^{NP}-C_{10}'^{NP}\right )\;, \hspace{2.0cm} A_S=0\;,
\end{eqnarray}
 where $C_{9,10}^{NP}$ and $C_{9,10}'^{NP}$  are the new Wilson coefficients arising due to leptoquark exchange and $E_K^*$ is the energy of the kaon in 
the $B$ meson rest frame and is given as
 \begin{equation}
 E_{K^*} = \frac{m^2_B + m^2_{K^*} - q^2}{2m_B}\;,
 \end{equation}
 The normalization constant $N$ is given as
 \begin{equation}
 N = \left[ \frac{G^2_F \alpha^2}{3 \cdotp2^{10}\pi^5 m_B}|V_{tb}V^*_{ts}|^2 \hat{s}\sqrt{\lambda}\beta_l\right]^{1/2}\;,
 \end{equation}
 where
 \begin{equation}
 \lambda = m^4_B + m^4_{K^*} +q^4 -2\left(m^2_B m^2_{K^*} + m^2_{K^*}q^2 + m^2_B q^2 \right)\;,~~~~~{\rm and}~~~~~\hat s= q^2/m_B^2\;.
 \end{equation}
 The transversity amplitude $A_t$ contains $\Delta_\parallel$ and negligible for massless lepton. It contributes only for
 $m_l \neq 0$. The light-cone distribution amplitude for $K^*$ is given by \cite{feldmann}
 \begin{equation}
 \Phi_{\bar{K}^*,a} = 6u\left(1-u\right)\lbrace 1 + a_1 \left(\bar{K}^*\right)_a C_1^{\left(3/2\right)}\left(2u-1\right) 
+ a_2 \left(\bar{K}^*\right)_a C_2^{\left(3/2\right)}\left(2u-1\right)\rbrace\;,
 \end{equation}
where the moments are
 \begin{equation}
  \lambda^{-1}_{B,+} = \int_{0}^{\infty} d\omega\frac{\Phi_{B,+}\left(\omega\right)}{\omega}\;, \hspace{3.9cm}
 \end{equation}
 \begin{equation}
 \lambda^{-1}_{B,-}\left(q^2\right) = \frac{e^{-q^2/\left(m_B\omega_0\right)}}{\omega_0}\left[-\textbf{E}\textbf{i}\left(q^2/m_B\omega_0\right) + i\pi\right].
 \end{equation}
 The detailed expression for the function $ \mathcal{T}_a \left(a = \perp, \parallel\right) $ at NLO in the QCDf framework is given in Appendix D.
 
\section{Transversity amplitudes in the low recoil}
The transversity amplitudes to leading order in $1/m_b$ at low recoil are given as
\beqa
A_\perp ^{L,R} & = & i\Bigg[ \left((C_9^{eff}+C_9^{NP}+C_9'^{NP}) \mp (C_{10}+C_{10}^{NP}+C_{10}'^{NP})\right) + \kappa \frac{2\hat{m}_b}{\hat{s}}C_7^{eff} \Bigg] f_\perp\;, \nn\\
A_\parallel ^{L,R} & = & -i\Bigg[ \left((C_9^{eff}+C_9^{NP}-C_9'^{NP}) \mp (C_{10}+C_{10}^{NP}-C_{10}'^{NP})\right) 
+ \kappa \frac{2\hat{m}_b}{\hat{s}}C_7^{eff} \Bigg] f_\parallel\;,\nn\\
A_0 ^{L,R} & = & -i\Bigg[ \left((C_9^{eff}+C_9^{NP}-C_9'^{NP}) \mp (C_{10}+C_{10}^{NP}-C_{10}'^{NP})\right) + \kappa \frac{2\hat{m}_b}{\hat{s}}C_7^{eff} \Bigg] f_0\;,
\eeqa
where the form factors read
\begin{eqnarray}
f_\perp = Nm_B\frac{\sqrt{2\hat{\lambda}}}{1+\hat{m}_{K^*}} V ,\hspace{4.4cm} \\ 
f_\parallel =  Nm_B \sqrt{2}\left(1+\hat{m}_{K^*}\right) A_1 , \hspace{3.3cm} \\
f_0 = Nm_B \frac{\left(1-\hat{s}-\hat{m}^2_{K^*}\right)\left(1+\hat{m}_{K^*}\right)^2 A_1
 - \hat{\lambda}A_2}{2 \hat{m}_{K^*}\left(1+\hat{m}_{K^*}\right)\sqrt{\hat{s}}}\;,
\end{eqnarray}
and the normalization factor is
\begin{equation}
N = \left[ \frac{G^2_F \alpha^2 |\lambda_t|^2 m_B \hat{s} \sqrt{\hat{\lambda}}}{3 \cdot2^{10}\pi^5}\right]^{1/2}\;.
\end{equation}
Here the dimensionless variables are $\hat{s} = q^2/m^2_B$ , $\hat{m}_i = m_i/m_B$ and $\hat{\lambda} = 
1+\hat{s}^2 +\hat{m}_{K^*}^4 - 2\left(\hat{s} + \hat{s}\hat{m}_{K^*}^2 + \hat{m}_{K^*}^2\right)$ and
the effective coefficients including the four-quark and gluon dipole operators are given by \cite{ref43}
\begin{equation}
C_7^{eff}=C_7-\frac{1}{3} \left[C_3+\frac{4}{3}C_4+20 C_5+\frac{80}{3} C_6 \right]+ 
\frac{\alpha_s}{4\pi}\left[\left(C_1-6 C_2\right)A(q^2)-C_8 F_8^{(7)}(q^2)\right]\;,
\end{equation}
\beqa
C_9^{eff}&=&C_9+h\left(0, q^2\right) \left[\frac{4}{3}C_1+C_2+\frac{11}{2}C_3 - \frac{2}{3}C_4 +52C_5 -\frac{32}{3}C_6\right]\nn\\
&-&\frac{1}{2}h\left(m_b, q^2\right)\left[7C_3+\frac{4}{3}C_4+76C_5+\frac{64}{3}C_6\right]+ \frac{4}{3}\left[C_3+\frac{16}{3}C_5
+\frac{16}{9}C_6\right]\nn\\
&+&\frac{\alpha_s}{4\pi}\left[C_1\left(B(q^2)+4C(q^2)\right)-3C_2\left(2B\left(q^2\right)-C\left(q^2\right)\right)-C_8F_8^{(9)}\left(q^2\right)\right]
\nn\\
& +& 8\frac{m^2_c}{q^2}\left[\left(\frac{4}{9}C_1+\frac{1}{3}C_2\right)(1+\lambda_u) +2C_3+20C_5\right]\;.
\eeqa
These include the CKM suppressions and the QCD matching corrections at next-to-leading order proportional to 
$\lambda_u = ({V_{ub}V^*_{us}})/{(V_{tb}V^*_{ts})}$, which corresponds to the small amount of CP-violation in the SM.

\section{$J_i$ coefficients}

In terms of the transversity amplitudes $A_0$, $A_\parallel$, $A_\perp$, and $A_t$ the $J_i$  coefficients can be expressed as \cite{ball, egede}
 \begin{equation}
 J^s_1 = \frac{\left(2+\beta ^2_l\right)}{4}\Bigg[|A_\perp ^L|^2 + |A_\parallel ^L|^2 + \left(L\rightarrow R\right)\Bigg] + \frac{4m^2_l}{q^2} 
{\rm Re}\left(A_\perp ^L A_\perp ^{R^*} + A_\parallel ^L A_\parallel ^{R^*}\right),
 \end{equation}
 \begin{equation}
  J^c_1 = |A_0^L|^2 + |A_0^R|^2 +\frac{4m^2_l}{q^2}\Bigg[|A_t|^2 + 2{\rm Re}\left(A_0^L A_0^{R^*}\right)\Bigg] + \beta^2_l |A_S|^2,\hspace{2.1cm}
 \end{equation}
 \begin{equation}
  J^s_2 = \frac{\beta^2_l}{4}\left[|A_\perp ^L|^2 + |A_\parallel ^L|^2 + \left(L\rightarrow R\right)\right],\hspace{6.3cm}
 \end{equation}
 \begin{equation}
  J^c_2 = -\beta^2_l\left[|A_0^L|^2 +\left(L\rightarrow R\right)\right],\hspace{7.6cm}
 \end{equation}
 \begin{equation}
 J_3 = \frac{1}{2}\beta^2_l\left[|A_\perp ^L|^2 - |A_\parallel ^L|^2 + \left(L \rightarrow R\right)\right],\hspace{6.2cm}
 \end{equation}
 \begin{equation}
 J_4 = \frac{1}{\sqrt{2}}\beta^2_l\left[{\rm Re}\left(A_0^L A_\parallel ^{L^*}\right) + \left(L \rightarrow R\right)\right],\hspace{6.1cm}
 \end{equation}
 \begin{equation}
 J_5 = \sqrt{2}\beta_l\left[{\rm Re}\left(A_0^L A_\perp ^{L^*}\right) - \left(L \rightarrow R\right) - 
\frac{m_l}{\sqrt{q^2}} {\rm Re}\left(A_\parallel ^L A^* _S + A_\parallel ^R A^* _S\right)\right],\hspace{1.1cm}
 \end{equation}
 \begin{equation}
  J^s_6 = 2\beta_l\left[{\rm Re}\left(A_\parallel ^L A_\perp ^{L^*}\right) - \left(L \rightarrow R\right)\right],\hspace{6.5cm}
 \end{equation}
 \begin{equation}
  J^c_6 = 4\beta_l\frac{m_l}{\sqrt{q^2}} {\rm Re}\left[A_0^L A_S^* + \left(L \rightarrow R\right)\right],\hspace{6.3cm}
 \end{equation}
 \begin{equation}
 J_7 = \sqrt{2}\beta_l\left[{\rm Im}\left(A_0^L A_\parallel ^{L^*}\right) - \left(L \rightarrow R\right) + \frac{m_l}{\sqrt{q^2}} 
{\rm Im}\left(A_\perp ^L A^* _S + A_\perp ^R A^* _S\right)\right],\hspace{1cm}
 \end{equation}
 \begin{equation}
  J_8 = \frac{1}{\sqrt{2}}\beta^2_l\left[{\rm Im}\left(A_0^L A_\perp ^{L^*}\right) + \left(L \rightarrow R\right)\right],\hspace{6.1cm}
 \end{equation}
 \begin{equation}
 J_9 = \beta^2_l\left[{\rm Im}\left(A_\parallel^{L^*} A_\perp ^L\right) + \left(L \rightarrow R\right)\right],\hspace{6.7cm}
 \end{equation}
 where 
 \begin{equation}
 \beta_l = \sqrt{1-\frac{4m^2_l}{q^2}}\;,
 \end{equation}
 and
 \begin{equation}
 A_i A_j^* = A_{i}^{ L}\left(q^2\right) A^{* L}_{j}\left(q^2\right) + A_{i}^{ R}\left(q^2\right) A^{* R}_{j}\left(q^2\right) 
\hspace{1cm} \left(i,j = 0, \parallel, \perp\right),
 \end{equation}
 in shorthand notation. The transversity amplitudes $A_i^{L,R}$ are presented in the appendix for A (B) for  low (high) $q^2$ 
region.
\section{ $\mathcal{T}_a^\pm$ calculation}
The  $B\rightarrow K^*$ matrix elements in large recoil limit depend on four independent functions $\mathcal{T}_a^\pm$ 
corresponding to a transversely $\left(a = \perp\right)$ and longitudinally $\left(a =\parallel\right)$ polarized $K^*$ and  
at next-to-leading order is given by \cite{feldmann}
\begin{equation}
\mathcal{T}_a = \xi_a C_a + \frac{\pi^2}{N_c}\frac{f_B f_{K^* ,a}}{m_B} \Xi_a \sum\limits_\pm \int \frac{d\omega}{\omega} 
\Phi_{B,\pm}\left(\omega\right) \int_{0}^{1} du \Phi_{K^*,a}\left(u\right) \mathcal{T}_{a,\pm}\left(u,\omega\right)\;,
\end{equation}
where $\Xi_\perp \equiv 1$, $\Xi_\parallel \equiv m_{K^*}/E_{K^*}$ and the factorization scale $\mu_f = \sqrt{m_b \Lambda_{QCD}}$.\\
The coefficient functions $C_a$ and $\mathcal{T}_{a , \pm}$ can be written as
\begin{equation}
C_a = C_a^{ (0)} +  \frac{\alpha_s\left(\mu_b\right) C_F}{4\pi}C_a^{ (1)}\;, \hspace{2cm}
\end{equation} 
and
 \begin{equation}
\mathcal{T}_{a , \pm} = \mathcal{T}_{a , \pm}^{(0)} \left(u ,\omega\right) + \frac{\alpha_s\left(\mu_f\right) C_F}{4\pi}
\mathcal{T}_{a , \pm}^{(1)} \left(u ,\omega\right)\;.
\end{equation}
 The form factor terms $C_a^{\left(0\right)}$ at leading order are
 \begin{equation}
 C_\perp^{ \left(0\right)} = C_7^{eff} + \frac{q^2}{2m_b m_B} Y\left(q^2\right)\;,
 \end{equation}
 \begin{equation}
 C_\parallel^{ \left(0\right)} = -C_7^{eff} - \frac{ m_B}{2m_b} Y\left(q^2\right)\;.
 \end{equation}
 The coefficients $C_a^{\left(1\right)}$ at next-to-leading order can be divided into a factorizable and a  non-factorizable part as
\begin{equation}
C_a^{\left(1\right)} = C_a^{ \left(f\right)} + C_a^{ \left(nf\right)}\;.
\end{equation}
 At NLO the factorizable correction reads
 \begin{equation}
 C_\perp^{ \left(f\right)} = C_7^{eff} \left(\ln \frac{m_b^2}{\mu^2}- L + \Delta M \right)\;,
 \end{equation}
\begin{equation}
 C_\parallel^{\left(f\right)} = -C_7^{eff} \left(\ln \frac{m_b^2}{\mu^2}+ 2L + \Delta M \right)\;,
\end{equation} 
and the non-factorizable correction for heavy to light transitions are
 \beqa
 C_F  C_\perp^{ \left(nf\right)} &= & -\bar{C}_2 F_2^{\left(7\right)} - C_8^{eff}F_8^{\left(7\right)} -\frac{q^2}{2m_b m_B} 
\left[\bar{C_2} F_2^{\left(9\right)} + 2\bar{C}_1\left( F_1^{\left(9\right)}+\frac{1}{6}F_2^{\left(9\right)}\right) 
+ C_8^{eff}F_8^{\left(9\right)}\right],\nn\\
  C_F  C_\parallel^{ \left(nf\right)}& = & \bar{C}_2 F_2^{\left(7\right)} + C_8^{eff}F_8^{\left(7\right)} 
+\frac{ m_B}{2m_b} \left[\bar{C_2} F_2^{\left(9\right)} + 2\bar{C}_1\left( F_1^{\left(9\right)}+\frac{1}{6}F_2^{\left(9\right)}\right) 
+ C_8^{eff}F_8^{\left(9\right)}\right]\;, 
 \eeqa
where $L$ and $\Delta M$ have given in Ref. \cite{feldmann}.
At leading order the hard-spectator scattering term $\mathcal{T}_{a , \pm}^{(0)} \left(u ,\omega\right)$ from weak annihilation diagram is given as
\begin{equation}
\mathcal{T}_{\perp , +}^{(0) } \left(u ,\omega\right) = \mathcal{T}_{\perp , -}^{(0) } \left(u ,\omega\right) = \mathcal{T}_{\parallel , +}^{(0) } 
\left(u ,\omega\right) = 0\;,
\end{equation}
\begin{equation}
\hspace{1cm} \mathcal{T}_{\parallel , -}^{(0) } \left(u ,\omega\right) = -e_q \frac{m_B \omega}
{m_B \omega -q^2 -i\epsilon}\frac{4m_B}{m_b} \left(\bar{C}_3 + 3\bar{C_4}\right)\;.
\end{equation}
The hard scattering functions $\mathcal{T}_a^{(1)}$ at next to leading order contain a factorisable as well as non-factorizable part 
\begin{equation}
\mathcal{T}_a^{(1)} = \mathcal{T}_a^{(f)} + \mathcal{T}_a^{(nf)}\;.
\end{equation}
Including $\mathcal{O}\left(\alpha_s\right)$ corrections the factorizable term to the hard scattering functions $\mathcal{T}_{a ,\pm}^{(1)}$ 
are given by
\begin{equation}
\mathcal{T}_{\perp ,+}^{(f) } \left(u, \omega\right) = C_7^{eff}\frac{2m_B}{\bar{u}E_{K^*}}\;,\hspace{1cm}
\end{equation}
\begin{equation}
\mathcal{T}_{\perp ,-}^{(f) } \left(u, \omega\right) = \mathcal{T}_{\parallel ,-}^{(f) } \left(u, \omega\right) = 0\;,
\end{equation}
\begin{equation}
\mathcal{T}_{\parallel ,+}^{(f) } \left(u, \omega\right) = C_7^{eff}\frac{4m_B}{\bar{u}E_{K^*}}\;,\hspace{1cm}
\end{equation}
and the non-factorizable correction can be computed by solving  the matrix elements of four-quark operators and the chromomagnetic dipole operator 
\beqa
\mathcal{T}_{\perp,+}^{(nf) } \left(u, \omega\right) & = & -\frac{4e_d C_8^{eff}}{u + \bar{u}q^2/m_B^2} + \frac{m_B}{2m_b}[ e_u t_\perp 
\left(u, m_c\right) \left(\bar{C_2} + \bar{C_4} - \bar{C_6}\right)\nn \\ 
& + & e_d t_\perp \left(u, m_b\right) \left(\bar{C_3} + \bar{C_4} - \bar{C_6}-4m_b/m_B \bar{C_5}\right) + e_d t_\perp \left(u, 0\right)\bar{C_3} ]\;, 
\eeqa
\begin{equation}
\mathcal{T}_{\perp,-}^{(nf) } \left(u, \omega\right) = 0\;,\hspace{9.8cm}
\end{equation}
\beqa
\mathcal{T}_{\parallel,+}^{(nf) } \left(u, \omega\right) & = & \frac{m_B}{m_b} [ e_u t_\parallel \left(u, m_c\right) \left(\bar{C_2}
 + \bar{C_4} - \bar{C_6}\right)\nn \\ & + & e_d t_\parallel \left(u, m_b\right) \left(\bar{C_3} + \bar{C_4} - \bar{C_6} \right) 
+ e_d t_\parallel \left(u, 0\right)\bar{C_3} ]\;,
\eeqa
\beqa
\mathcal{T}_{\parallel,-}^{(nf) } \left(u, \omega\right) & = & e_q \frac{m_B\omega}{m_B\omega - q^2 - i\epsilon} \Bigg[ 
\frac{8C_8^{eff}}{\bar{u} + uq^2/m_B^2}\nn \\& + & \frac{6m_B}{m_b} \Bigg(h\left(\bar{u}m_B^2 + uq^2 , m_c\right)\left(\bar{C}_2 + 
\bar{C}_4 + \bar{C}_6\right)\nn \\ & +& h\left(\bar{u}m_B^2 + uq^2 , m_b^{pole}\right)\left(\bar{C}_3 + \bar{C}_4 + \bar{C}_6\right)\nn \\ 
& + &h\left(\bar{u}m_B^2 + uq^2 , 0\right)\left(\bar{C}_3 + 3\bar{C}_4 + 3\bar{C}_6\right)-\frac{8}{27} \left(\bar{C}_3 - 
\bar{C}_5 - 15\bar{C}_6\right)\Bigg) \Bigg].
\eeqa
The $t_a \left(u ,m_q\right)$ functions are given by
\beqa
t_\perp \left(u ,m_q\right) &=& \frac{2m_B}{\bar{u}E_{K^*}}I_1\left(m_q\right) + \frac{q^2}{\bar{u}^2E^2_{K^*}}\left(B_0\left(\bar{u}m_B^2 + uq^2 , m_q \right)
 - B_0\left(q^2 , m_q\right)\right),\nn\\
t_\parallel \left(u ,m_q\right)& = & \frac{2m_B}{\bar{u}E_{K^*}}I_1\left(m_q\right) + \frac{\bar{u}m_B^2 + uq^2}{\bar{u}^2E^2_{K^*}}
\left(B_0\left(\bar{u}m_B^2 + uq^2 , m_q \right) - B_0\left(q^2 , m_q\right)\right)\;,
\eeqa
where $B_0$ and $I_1$ are 
\begin{equation}
B_0\left(q^2 ,m_q\right) = -2\sqrt{4m^2_q/q^2 - 1} \arctan \frac{1}{\sqrt{4m^2_q/q^2 - 1}}\;,\hspace{2.3cm}
\end{equation}
\begin{equation}
I_1\left(m_q\right) = 1 + \frac{2m^2_q}{\bar{u}\left(m^2_B - q^2\right)}\left[L_1\left(x_+\right) + L_1\left(x_-\right) - L_1\left(y_+\right) 
- L_1\left(y_-\right)\right]\;,
\end{equation}
and 
\begin{equation}
x_\pm = \frac{1}{2} \pm \left(\frac{1}{4} - \frac{m^2_q}{\bar{u}m_B^2 + uq^2}\right) ^{1/2}, \hspace{0.8cm} y_\pm
 = \frac{1}{2} \pm \left(\frac{1}{4} - \frac{m^2_q}{q^2}\right)^{1/2}\;, 
\end{equation}
\begin{equation}
L_1\left(x\right) = \ln \frac{x-1}{x} \ln\left(1-x\right) - \frac{\pi^2}{6} + Li_2\left(\frac{x}{x-1}\right)\;. \hspace{2.3cm}
\end{equation}
\section{Functions involved in Isospin asymmetry parameter}
The function $K_1^\parallel$ receives an annihilation contribution to leading order and the function $K_{1,2}^\perp$ appear 
at subleading order of the $\Lambda_h/m_B$ expansion. The meson photon transition form factors and their role in the annihilation contribution to 
$B$ meson decays has been given by the function with superscript (a). Here only transverse polarization contributes and 
there are no  contributions from $C_5$ and $C_6$. The function $K_{1,2}^{(b)}$ contain the decay amplitude from the diagram of hard spectator 
interactions involving the gluonic penguin operator $\mathcal{O}_8$ and the contributions of the hard spectator interactions diagrams involving 
the operator $\mathcal{O}_{1-6}$ have been included in $K_{1,2}^{(b)}$. \\
The functions $K^\parallel_1$ and $K^\perp_{1,2}$ defined via $b^a_q$ is given as \cite{lord}
\begin{equation}
K_1^\perp \left(q^2\right) = K_1^{\perp \left(a\right)} \left(q^2\right) + K_1^{\perp \left(b\right)} \left(q^2\right) 
+ K_1^{\perp \left(c\right)} \left(q^2\right)\;, \hspace{4cm}
\end{equation}
with
\begin{eqnarray}
&&K_1^{\perp \left(a\right)} \left(q^2\right) = -\left(\bar{C}_6\left(\mu_h\right) + \frac{\bar{C}_5\left(\mu_h\right)}{N_c}\right) F_\perp\left(\hat{s}\right) ,
\hspace{0.8cm} F_\perp\left(\hat{s}\right) = \frac{1}{3} \int_{0}^{1} du \frac{\phi_ {K^*}^\perp \left(u\right)}{\bar{u} + u\hat{s}},\nn\\
&&K_1^{\perp \left(b\right)} \left(q^2\right) = C^{eff}_8\left(\mu_h\right) \frac{m_b}{m_B} \frac{C_F}{N_c} 
\frac{\alpha_s\left(\mu_h\right)}{4\pi} X_\perp \left(\hat{s}\right) ,~~~
 X_\perp \left(\hat{s}\right)  = F_\perp \left(\hat{s}\right) + \frac{1}{3} \int_{0}^{1} du \frac{\phi_{K^*}^\perp\left(u\right)}{(\bar{u} + u\hat{s})^2},\nn\\
&&K_1^{\perp \left(c\right)} \left(q^2\right) = \frac{C_F}{N_c} \frac{\alpha_s\left(\mu_h\right)}{4\pi}\frac{2}{3} \int_{0}^{1} du 
\frac{\phi_{K^*}^\perp \left(u\right)}{\bar{u} + u\hat{s}} F_V\left(\bar{u}m_B^2 + uq^2\right)
\end{eqnarray}
and
\begin{equation}
K_2^\perp \left(q^2\right) = K_2^{\perp \left(a\right)} \left(q^2\right) + K_2^{\perp \left(b\right)} \left(q^2\right) 
+ K_2^{\perp \left(c\right)} \left(q^2\right)\;, \hspace{4cm}
\end{equation}
with
\begin{eqnarray}
&&K_2^{\perp \left(a\right)} \left(q^2\right) = -\frac{\lambda_u}{\lambda_t}\left(\frac{\bar{C}_1}{3}\left(\mu_h\right) 
+ \bar{C}_2 \left(\mu_h\right)\right)\delta_{qu} + \left( \bar{C}_4 \left(\mu_h\right) + \frac{\bar{C}_3 \left(\mu_h\right)}{3} \right)\;,\nn\\
&& 
K_2^{\perp \left(b\right)} \left(q^2\right) = \mathcal{O}\left(\frac{\Lambda_h}{m_B}\right)\;,\nn\\
&&
K_2^{\perp \left(c\right)} \left(q^2\right) = -\frac{C_F}{N_c} \frac{\alpha_s\left(\mu_h\right)}{4\pi} \frac{1}{2} \int_{0}^{1} du 
\left(g_\perp^{(\nu)}\left(u\right) - \frac{g_\perp^{' (a)}\left(u\right)}{4}\right) F_V\left(\bar{u}m_B^2 + uq^2\right)\;.
\end{eqnarray}
For parallel case in Eq. (61),
\begin{equation}
K_1^\parallel \left(q^2\right) = K_1^{\parallel \left(a\right)} \left(q^2\right) + K_1^{\parallel \left(b\right)} \left(q^2\right) 
+ K_1^{\parallel \left(c\right)} \left(q^2\right)\;, \hspace{3cm}
\end{equation}
with
\begin{eqnarray}
&& K_1^{\parallel \left(a\right)} \left(q^2\right) = K_2^{\perp \left(a\right)} \left(q^2\right),\nn\\
&& K_1^{\parallel \left(b\right)} \left(q^2\right) = -C^{eff}_8\left(\mu_h\right) \frac{m_b}{m_B} \frac{C_F}{N_c} 
\frac{\alpha_s\left(\mu_h\right)}{4\pi} F_\parallel\left(\hat{s}\right) ,\hspace{0.8cm} F_\parallel\left(\hat{s}\right) = 
2\int_{0}^{1} du \frac{\phi_\parallel\left(u\right)}{\bar{u} + u\hat{s}},\nn\\
&&K_1^{\parallel \left(c\right)} \left(q^2\right) = -\frac{C_F}{N_c} \frac{\alpha_s\left(\mu_h\right)}{4\pi}2\int_{0}^{1} du 
\phi_\parallel\left(u\right) F_V\left(\bar{u}m_B^2 + uq^2\right)\;.
\end{eqnarray}
The vector form factor $F_V\left(s\right)$ is given by 
\begin{eqnarray}
F_V\left(s\right) &= & \frac{3}{4} \lbrace h\left(s ,m_c ,\mu_h \right) \left(\bar{C}_2\left(\mu_h\right) + \bar{C}_4\left(\mu_h\right)
 + \bar{C}_6\left(\mu_h\right)\right)\nn \\ & +& h\left(s ,m_b ,\mu_h \right) \left(\bar{C}_3\left(\mu_h\right) + \bar{C}_4\left(\mu_h\right)
 + \bar{C}_6\left(\mu_h\right)\right)\nn \\ & + & h\left(s ,0 ,\mu_h \right)\left(\bar{C}_3\left(\mu_h\right) + 3\bar{C}_4\left(\mu_h\right) 
+ 3\bar{C}_6\left(\mu_h\right)\right) \nn\\ &-&\frac{8}{27} \left(\bar{C}_3\left(\mu_h\right) - \bar{C}_5\left(\mu_h\right) -
 15 \bar{C}_6\left(\mu_h\right)\right)\rbrace\;.
\end{eqnarray}

\end{document}